\documentclass[journal,draftcls,onecolumn,letterpaper,twoside,11pt]{IEEEtran}

\usepackage[cmex10]{amsmath}
\usepackage{latexsym,amssymb,cite,bbm,epsf,amsthm,subfigure,algorithm,algorithmic}
\usepackage{graphicx}
\usepackage{enumerate}
\usepackage{mathrsfs}
\usepackage{epstopdf}
\usepackage{bbm}

\newcommand{\bc}{\begin{center}}
\newcommand{\ec}{\end{center}}
\newcommand{\be}{\begin{equation}}
\newcommand{\ee}{\end{equation}}
\newcommand{\ba}{\begin{array}}
\newcommand{\ea}{\end{array}}
\newcommand{\bea}{\begin{eqnarray}}
\newcommand{\eea}{\end{eqnarray}}
\newcommand{\bal}{\begin{align}}
\newcommand{\eal}{\end{align}}
\newcommand{\ei}{\end{itemize}}
\newcommand{\bi}{\begin{itemize}}
\newcommand{\bfi}{\begin{figure}}
\newcommand{\efi}{\end{figure}}
\newcommand{\MB}{\left[\begin{array}}
\newcommand{\ME}{\end{array}\right]}
\newcommand{\nn}{\nonumber}

\newtheorem{thm}{Theorem}
\newtheorem*{dfn}{Definition}

\newtheorem{rem}{Remark}

\renewcommand{\vec}[1]{\mbox{\boldmath${#1}$}}

\newcommand{\Exp}{\mathsf{E}}

\newcommand{\Pro}{\mathsf{P}}
\newcommand{\Hyp}{\mathsf{H}}

\newcommand{\cT}{\mathcal{T}}

\newcommand{\cN}{\mathcal{N}}

\newcommand{\bN}{\mathbb{N}}
\newcommand{\bR}{\mathbb{R}}

\newcommand{\vh}{\vec{h}}
\newcommand{\vy}{\vec{y}}
\newcommand{\va}{\vec{a}}
\newcommand{\mG}{\vec{G}}

\newcommand{\ignore}[1]{{}}

\usepackage{tikz}

\begin{document}

\title{Sequential Distributed Detection in Energy-Constrained Wireless Sensor Networks}

\author{Yasin~Yilmaz\IEEEauthorrefmark{1}\footnote{\IEEEauthorrefmark{2}Electrical Engineering Department, Columbia University, New York, NY 10027.}\;\;
        and \, Xiaodong Wang\IEEEauthorrefmark{1}}

\maketitle

\begin{abstract}
The recently proposed sequential distributed detector based on level-triggered sampling operates as simple as the decision fusion techniques and at the same time performs as well as the data fusion techniques. Hence, it is well suited for resource-constrained wireless sensor networks. However, in practical cases where sensors observe discrete-time signals, the random overshoot above or below the sampling thresholds considerably degrades the performance of the considered detector. We propose, for systems with stringent energy constraints, a novel approach to tackle this problem by encoding the overshoot into the time delay between the sampling time and the transmission time. Specifically, each sensor computes the local log-likelihood ratio (LLR) and samples it using level-triggered sampling. Then, it transmits a single pulse to the fusion center (FC) after a transmission delay that is proportional to the overshoot, as in pulse position modulation (PPM). The FC, upon receiving a bit decodes the corresponding overshoot and recovers the transmitted LLR value. It then updates the approximate global LLR and compares it with two threshold to either make a decision or to continue the sequential process. We analyze the asymptotic average detection delay performance of the proposed scheme.
We then apply the proposed sequential scheme to target detection in wireless sensor networks under the four Swerling fluctuating target models. It is seen that the proposed sequential distributed detector offers significant performance advantage over conventional decision fusion techniques.
\end{abstract}


\section{Introduction}
\label{sec:intro}

We consider the problem of distributed detection where a number of sensors, under energy constraints, communicate to a fusion center (FC) which is responsible for making the final decision.
In \cite{Tenney81} it was shown that under a fixed fusion rule, with two sensors each transmitting one bit information to the FC, the optimum local decision rule is a likelihood ratio test (LRT) under the Bayesian criterion.
Later, in \cite{Chair86} and \cite{Thomo87} it was shown that the optimum fusion rule at the FC is also an LRT under the Bayesian and the Neyman-Pearson criteria, respectively.
It was further shown in \cite{Tsitsiklis88} that as the number of sensors tends to infinity it is asymptotically optimal to have all sensors perform an identical LRT.
The case where sensors observe correlated signals was also considered, e.g., \cite{Aalo89}.

Most works on distributed detection, including the above mentioned, treat the fixed-sample-size approach where each sensor collects a fixed number of samples and the FC makes its final decision at a fixed time. There is also a significant volume of literature that considers the sequential distributed detection, e.g., \cite{Veer93,Hussain94,Fellouris11,Yilmaz12,Yilmaz13}.
The sequential probability ratio test (SPRT) is the optimum sequential (centralized, i.e., non-distributed) test for i.i.d. observations in terms of minimizing the average sample number (detection delay) among all sequential tests satisfying the same
error probability constraints \cite{Wald48}. The SPRT has been shown in \cite[Page 109]{Poor} to asymptotically require, on average, four times less samples (for Gaussian signals) to reach a decision than the
best fixed-sample-size test, for the same level of confidence. The distributed schemes in \cite{Fellouris11,Yilmaz12,Yilmaz13} follow the SPRT procedure at the FC by reporting the local test statistics from sensors to the FC via \emph{level-triggered sampling}, a nonuniform sampling technique in which sampling times are dynamically determined by the signal to be sampled. Such a sampling scheme naturally outputs low-rate information
(e.g., one bit per sample) without performing any quantization, which is ideally suited for applications with stringent energy constraints.
Data fusion (multi-bit messaging) is known to be much more powerful than decision fusion (one-bit messaging) \cite{Chaud12}, albeit it consumes higher energy. Moreover, the recently proposed sequential detection schemes  based on level-triggered sampling in \cite{Fellouris11,Yilmaz12,Yilmaz13} are as powerful as data-fusion techniques, and at the same time they are as simple and energy-efficient as decision-fusion techniques.

In a practical system where sensors observe discrete-time signals, the major problem in the level-triggered sampling procedure is the random overshoots above or below the sampling thresholds. In the overshoot-free case, where sensors observe continuous-time signals with continuous paths, the sequential distributed detector based on level-triggered sampling, using a single bit per sample, achieves the order-2 asymptotic optimality \footnote{The definition of asymptotic optimality is given in Theorem \ref{thm:finite}. Order-2 implies order-1, which is a weaker type and the most frequent form of asymptotic optimality encountered in the literature.}, which can be seen as the best performance for sequential distributed detectors. It is not possible to inform the FC about the overshoot value using the one-bit information that is the natural output of the sampling mechanism at the corresponding sampling time. To tackle the overshoot problem, in \cite{Fellouris11} and \cite{Yilmaz13}, the log-likelihood ratio of the bit received by the FC is computed via simulations, which includes an average value for the overshoot. Alternatively in \cite{Yilmaz12} the overshoot in each sample is reported to the FC at the sampling time in a quantized form, resulting in a multi-bit scheme. The former approach achieves the order-1 asymptotic optimality, whereas the latter achieves the order-2 asymptotic optimality if the number of quantization bits increases at a reasonably low rate with the decreasing target error probabilities. In other words, to achieve small target error probabilities, we need to transmit multiple bits of information.

Ultra-wideband (UWB) communications, in which sensors transmit short pulses with low power, suits well to energy-constrained sensor networks \cite{UWB06}. In UWB, pulse position modulation (PPM), which encodes information in time, is among the most popular modulation schemes \cite{Aiello03}.
In this paper, we accordingly propose to encode overshoot into the time-delay between sampling time and transmission time. We show that the result of \cite{Yilmaz12} on the order-2 asymptotic optimality holds for the proposed scheme. Specifically, the sequential distributed detector in this paper achieves the order-2 asymptotic optimality if the available bandwidth increases at a certain rate with the decreasing target error probabilities. Note that the technique proposed to encode overshoot matches well with the level-triggered sampling mechanism since it also encodes information in sampling times, as opposed to the traditional uniform-in-time sampling, which encodes information only in the sampled values. Accordingly, the level-triggered sampling mechanism is called a time encoding machine in neural networks \cite[Section II-C]{Lazar11}. As an application of the proposed sequential distributed detector, we consider target detection in wireless sensor networks.

The contributions of this work are twofold: (i) an energy-efficient and asymptotically optimal sequential distributed detector which transmits a single pulse for each sample from sensors to the FC by encoding the random overshoot in each sample into the time delay between sampling time and transmission time, (ii) application to target detection in wireless sensor networks under the four Swerling target models.

The remainder of the paper is organized as follows. In Section II, after formulating the problem, we summarize the sequential distributed detector of interest, which is based on level-triggered sampling. The proposed approach to tackling the overshoot problem is given and analyzed in Section III. In Section IV, we apply the proposed scheme to target detection in wireless sensor networks. Finally the paper is concluded in Section V.

\section{Sequential Distributed Detection via Level-triggered Sampling }
\label{sec:info}

Consider a wireless sensor network with $K$ sensors and a fusion center (FC) which is responsible for making the final decision. Each sensor $k$ observes a discrete-time signal $y_t^k,~t\in\bN$, which it reports to the FC in some form.
Assume that the observations at each sensor are i.i.d. and the sensors are independent.
Denote $f_i^k,~i=0,1$ as the pdf of the signal observed by sensor $k$ under hypothesis $\Hyp_i$. Then, the log-likelihood ratio (LLR) of the single observation $y_m^k$, the local LLR at sensor $k$ up to time $t$, and the global LLR at the FC up to time $t$ are given respectively by
\be
\label{eq:LLR3}
    l_{m}^k \triangleq \log \frac{f_1^k(y_{m}^k)}{f_0^k(y_{m}^k)},~~ L_t^k = \sum_{m=1}^t l_{m}^k,~~ \text{and} ~~
    L_t = \sum_{k=1}^K L_t^k.
\ee

In distributed detection, each sensor $k$ needs to sample its local LLR $L_t^k$ and transmit a few times for each sampled value to the FC.
In particular, in the level-triggered sampling \cite{Fellouris11,Yilmaz12}, the local LLR $L_t^k$ is sampled at a sequence of random times $\{t_n^k\}_n$ that are dynamically determined by $L_t^k$ itself. Specifically, the $n$th sample is taken at time $t_n^k$ when the change in the signal since the last sampling time $t_{n-1}^k$ exceeds a predetermined threshold $\Delta$, i.e.,
\be
\label{eq:samp_time}
	t_n^k \triangleq \min\{ t\in\bN: |L_t^k-L_{t_{n-1}^k}^k| \geq \Delta  \},~L_0^k=0,
\ee
where $\Delta$ is determined using
\be
\label{eq:delta}
    \Delta \tanh\left(\frac{\Delta}{2}\right)=\frac{1}{R}\sum_{k=1}^K |\Exp_i[L_1^k]|
\ee
for the FC to receive messages with an average rate of $R$ messages per unit time interval under $\Hyp_i$ \cite[Section IV-B]{Yilmaz12}. In practice, $\Delta$ can be set using \eqref{eq:delta} to satisfy a maximum or minimum average message rate constraint.
In \eqref{eq:delta} and throughout the paper, $\Exp_i[\cdot]$ denotes the expectation under hypothesis $\Hyp_i$.
The sample summary
\be
\label{eq:loc_dec}
    b_n^k \triangleq \text{sign}(\lambda_n^k),
\ee
is a one-bit encoding of the change $\lambda_n^k \triangleq L_{t_n^k}^k-L_{t_{n-1}^k}^k$ in the LLR signal $L_t^k$ during the time interval $(t_{n-1}^k,t_n^k]$. In other words, $b_n^k$ represents which threshold ($\Delta$ or $-\Delta$) $\lambda_n^k$ exceeds. However, it does not represent how much $\lambda_n^k$ exceeds the threshold. Define $q_n^k \triangleq |\lambda_n^k|-\Delta$ as the excess amount of LLR over $\Delta$ or under $-\Delta$, i.e., overshoot. Then, $\lambda_n^k$ is given by
\be
\label{eq:change_enc}
    \lambda_n^k=b_n^k(\Delta+q_n^k).
\ee

In \cite{Fellouris11}, sensor $k$ at time $t_n^k$ transmits $b_n^k$ to the FC, which approximates $\lambda_n^k$ by computing the LLR $\hat{\lambda}_n^k$ of $b_n^k$. Whereas in \cite{Yilmaz12}, at time $t_n^k$, in addition to $b_n^k$, sensor $k$ also transmits some additional bits quantizing the overshoot $q_n^k$, and then the FC computes $\hat{q}_n^k$ and accordingly, from \eqref{eq:change_enc}, an approximate value $\hat{\lambda}_n^k$ of $\lambda_n^k$ using the received bits.
Since we have
\be
\label{eq:LLR_recover}
	L_{t_n^k}^k = \sum_{m=1}^n \left( L_{t_m^k}^k-L_{t_{m-1}^k}^k \right) = \sum_{m=1}^n \lambda_m^k,
\ee
the approximate LLR messages $\{\hat{\lambda}_n^k\}_{n,k}$ are combined at the FC to compute the approximations $\hat{L}_t^k$ and $\hat{L}_t$ to the local and global LLRs $L_t^k$ and $L_t$, respectively, as follows,
\be
\label{eq:FC_LLR}
    \hat{L}_t^k = \sum_{m=1}^n \hat{\lambda}_m^k,~~ t \in [t_n^k,t_{n+1}^k),~ t_0^k=0,~~~ \text{and}~~~ \hat{L}_t = \sum_{k=1}^K \hat{L}_t^k.
\ee
In fact, the FC computes only $\hat{L}_t$ in a recursive way. Specifically, upon receiving the $n$th message in the global order at time $t_n$ from sensor $k_n$ the FC updates $\hat{L}_t$ as
\be
\label{eq:LLR_recur}
	\hat{L}_{t_n} = \hat{L}_{t_{n-1}} + \hat{\lambda}_n.
\ee
The approximate LLR signal $\hat{L}_t$ is kept constant until the arrival of the next message. In addition to updating the global LLR, the FC also runs a sequential test similar to SPRT. Specifically, at time $t_n$, $\hat L_{t_n}$ is compared with two thresholds, $-\hat{B}$ and $\hat{A}$, where $\hat{A},\hat{B}>0$. If $\hat L_{t_n} \in (-\hat{B},\hat{A})$, the test continues, otherwise it stops. At the stopping time
\be
\label{eq:FC_stop}
\hat{\cT}\triangleq \min\{t\in\bN: \hat{L}_t \not\in (-\hat{B},\hat{A})\},
\ee
the FC declares $\Hyp_0$ if $\hat L_{\hat{\cT}} \leq -\hat{B}$ and $\Hyp_1$ if $\hat L_{\hat{\cT}} \geq \hat{A}$, i.e.,
\be
\label{eq:FC_dec}
    \hat{\delta}_{\hat{\cT}} \triangleq \left\{ \ba{ll} 0 & \text{if}~\hat{L}_{\hat{\cT}} \leq -\hat{B}, \\ 1 & \text{if}~\hat{L}_{\hat{\cT}} \geq \hat{A}. \ea \right.
\ee
The thresholds $\hat{A}$ and $-\hat{B}$ are selected to meet the target false alarm and mis-detection probabilities, i.e., $\Pro_0(\hat{\delta}_{\hat{\cT}}=1)=\alpha$, $\Pro_1(\hat{\delta}_{\hat{\cT}}=0)=\beta$, where $\Pro_i$, $i=0,1$, denotes the probability measure under $\Hyp_i$.

\section{Proposed Detector with Time-encoded Overshoot}
\label{sec:time_enc}

The random overshoot $q_n^k$ causes a significant problem in recovering the LLR message $\lambda_n^k$ at the FC especially when $q_n^k$ takes values that are comparable to $\Delta$ [cf. \eqref{eq:change_enc}].
$q_n^k$ represents the missing LLR information at the FC, which is not encoded in $b_n^k$.
With $q_n^k$ not available at the FC the discrepancy between $\lambda_n^k$ and $\hat{\lambda}_n^k$ accumulates over time through the recursion in \eqref{eq:LLR_recur}, which in turn causes a significant performance gap between the distributed detector and the centralized one.
Existing works followed different approaches to overcome this problem. In \cite{Fellouris11}, the LLR of $b_n^k$ is computed via simulations, which in fact includes an average value for $q_n^k$. At each sampling time $t_n^k$ this average value is used to replace the unknown overshoot $q_n^k$ in \eqref{eq:change_enc}. This approach achieves the order-1 asymptotic optimality, but not order-2, which is a stronger type of asymptotic optimality. In \cite{Yilmaz12}, at each time $t_n^k$ some additional bits are used to quantize the overshoot $q_n^k$. This approach can achieve the order-2 asymptotic optimality if the number of quantization bits increases at a rate of $\log |\log \gamma|$ where $\gamma\to0$ at least as fast as the error probabilities $\alpha$ and $\beta$, i.e., $\gamma=O(\alpha), \gamma=O(\beta)$.

In this paper, we propose to encode $q_n^k$ in time while sending the sign bit $b_n^k$ to the FC, as in PPM in a UWB system, which is an ideal fit for energy-constrained sensor networks \cite{UWB06}.
In the proposed scheme, each sensor $k$ sends $b_n^k$ to the FC at time $\tau_n^k\in[t_n^k,t_n^k+1)$.
\ignore{
For completeness, we initially analyze the ideal case with infinite time resolution at sensors, i.e., $\tau_n^k\in\bR^+$, and the FC, which can perfectly recover $\tau_n^k$ through an infinite-bandwidth error-free communication link. This ideal system will serve as a benchmark for practical systems that we will develop later in this section.

\subsection{Infinite time resolution case}
\label{sec:infinite}

Assume for now that the FC, knowing $t_n^k$ and $\tau_n^k$, somehow computes $q_n^k$. We will describe how it computes later in this section. Then, upon receiving $b_n^k$ at time $\tau_n^k$ it recovers $\lambda_n^k$ and computes $L_{t_n^k}^k$ using \eqref{eq:change_enc} and \eqref{eq:LLR_recover}, respectively.
In other words, the local LLR $L_{t_n^k}^k$ is available to the FC at time $\tau_n^k$ with the delay $\xi_n^k \triangleq \tau_n^k-t_n^k \in [0,1),\forall k,n$. In our scheme, the FC, similarly to \eqref{eq:FC_LLR}, approximates the local and global LLR signals $L_t^k$ and $L_t$ with the staircase signals $\hat{L}_t^k$ and $\hat{L}_t$, respectively as follows
\begin{align}
\label{eq:FC_approx}
\begin{split}
    \hat{L}_t^k =& L_{t_n^k}^k,~~t\in[\tau_n^k,\tau_{n+1}^k),~~t_0^k=\tau_0^k=0,\\
    \text{and}~~\hat{L}_t =& \sum_{k=1}^K \hat{L}_t^k.
\end{split}
\end{align}
In fact, the FC employs the following procedure. Upon receiving the $n$th bit $b_n$ from any sensor at time $\tau_n$ (with a short delay $\xi_n$ after $b_n$ is actually sampled at time $t_n$) it computes $\lambda_n$ as in \eqref{eq:change_enc} (assuming the overshoot $q_n$ is computable by knowing $t_n, \tau_n$) and updates the approximate global LLR $\hat{L}_t$ as
\be
\label{eq:FC_update}
    \hat{L}_{\tau_n} = \hat{L}_{\tau_{n-1}} + \lambda_n,
\ee
which is kept constant until the arrival of the next bit $b_{n+1}$ from any sensor unless the procedure terminates at time $\tau_n$. The FC then performs the sequential test given in \eqref{eq:FC_stop} and \eqref{eq:FC_dec}.

Note from \eqref{eq:FC_approx} that the approximate local LLR $\hat{L}_t^k$ attains the exact local LLR $L_{t_n^k}^k$ at time $\tau_n^k$, shortly after the sampling time $t_n^k$, i.e., $\hat{L}_{\tau_n^k}^k=L_{t_n^k}^k$. This is in general not true for the approximate global LLR $\hat{L}_t$, i.e., $\hat{L}_{\tau_n^k}\not=L_{t_n^k}$, since $\hat{L}_{\tau_n^k}^j \not= L_{t_n^k}^j,~j\not=k$, for the other $k-1$ sensors. Next, we explain how the FC computes the overshoot $q_n^k$ by knowing $t_n^k$ and $\tau_n^k$.

We here propose to linearly encode the overshoot in time. Specifically, we use a linear function
\be
\label{eq:enc_fnc}
    g(q_n^k)=\xi_n^k=\frac{q_n^k}{r}
\ee
to encode the overshoot $q_n^k$ into the transmission delay $\xi_n^k$. The slope parameter $r$ of the linear function is known to sensors and the FC. The output of the encoding function is the transmission delay $\xi_n^k=\tau_n^k-t_n^k$, that is, the delay between the transmission time $\tau_n^k$ and the sampling time $t_n^k$. Assuming a global clock, i.e., $t\in\bN$ is common for all sensors and the FC, the FC, upon receiving $b_n^k$ at time $\tau_n^k$, can obtain $\xi_n^k\in[0,1)$, as the fractional part of $\tau_n^k$, i.e.,
$\xi_n^k = \tau_n^k - \lfloor \tau_n^k \rfloor$. Then, the overshoot can be recovered by $q_n^k=r\xi_n^k$, as shown in Fig. \ref{fig:overshoot}.

\begin{figure}[t]
\centering
\includegraphics[scale=0.7]{encoding.eps} ~~~
\includegraphics[scale=0.4]{time-enc_over.eps}
\caption{The encoding mechanism and the linear decoding function $g^{-1}$ with slope $r>\theta$, which maps the transmission delay $\xi_n^k$ to the overshoot $q_n^k=r\xi_n^k$.}
\label{fig:overshoot}
\end{figure}

To this end, we made the following assumptions.
\bi
\item[] \textbf{(A1)} There exists a global clock running in the wireless sensor network, hence the FC knows the potential sampling instants $t \in \bN$.
\item[] \textbf{(A2)} There exists a bound $\theta$ for overshoots and we set $r>\theta \triangleq \max_{k,n} q_n^k$. Then, each transmission delay $\xi_n^k\in[0,1),\forall k,n$.
\item[] \textbf{(A3)} Each pair of transmission time $\tau_n^k\in\bR^+$ and transmitted bit $b_n^k$ is available to the FC through ideal channels.
\ei
With (A1), (A2) and (A3), upon the FC receives $b_n^k$ at time instant $\tau_n^k$, the sampling time and the delay can be uniquely determined as $t_n^k= \lfloor \tau_n^k \rfloor$ and
$\xi_n^k = \tau_n^k - t_n^k$, respectively. If (A1) is not met, i.e., the network lacks a global clock, then a synchronization pulse (in addition to $b_n^k$) can be used to report each sampling time $t_n^k$, resulting in a two-pulse scheme. Note that (A2) defines a lower bound for $r$. The slope of the inverse encoding (decoding) function $g^{-1}$ must be larger than the maximal overshoot value $\theta$, as shown in Fig. \ref{fig:overshoot}. We do not have an upper bound for $r$, and in fact by choosing $r$ arbitrarily large we can have the transmission delay $\xi_n^k$ arbitrarily small. That is, the selection of the $r$ value strongly depends on the dynamic range of overshoots, i.e., $\theta$, and the unit time interval.

Note that the sensors observe and sample the local signals at time $t \in \bN$ under the global clock, but they transmit and the FC receives bits at time $\tau \in \bR^+$ after a delay $\xi\in[0,1)$, whereas in \cite{Fellouris11,Yilmaz12} the transmitting time and the sampling time are the same. We next analyze the asymptotic average detection delay performance of the proposed detector with infinite time resolution.

\begin{rem}
\label{rem:inf}
  The sequential distributed detector with infinite time resolution, given in \eqref{eq:FC_stop} -- \eqref{eq:enc_fnc}, is order-2 asymptotically optimal, i.e.,
  \be
  \label{eq:thm}
    \Exp_i[\hat{\cT}]-\Exp_i[\cT]=O(1),~i=0,1,~ \text{as}~ \alpha,\beta\to0,\nn
  \ee
  where $\cT = \min\{t\in\bN: L_t \not\in (-B,A)\}$ and $\hat{\cT}=\min\{t\in\bN: \hat{L}_t \not\in (-\hat{B},\hat{A})\}$ are the stopping times of the centralized SPRT, i.e., the optimum sequential detector, and the proposed detector, respectively, both of which satisfy the false alarm probability $\alpha$ and mis-detection probability $\beta$; and $O(1)$ denotes a constant term.
\end{rem}

This result is not surprising since, with the assumptions (A1) -- (A3), the random overshoot in each sample becomes available to the FC within a unit time interval. As a result, the order-2 asymptotic optimality result given in \cite{Fellouris11} for the no-overshoot case, in which sensors observe continuous-time signals with continuous paths, holds here. We next propose an energy-efficient sequential distributed detector for practical systems with finite time resolution, and analyze its asymptotic average detection delay performance.

\subsection{Finite time resolution case}
\label{sec:finite}
}
In a practical system, due to bandwidth and hardware limitations, we cannot transmit with an infinite time resolution, hence the transmission delay $\xi_n^k$ (resp. the transmission time $\tau_n^k$) takes a discrete set of values in $[0,1)$ (resp. $[t_n^k,t_n^k+1)$). Obviously, we can encode only a discrete set of $q_n^k$ values in $\xi_n^k$, resulting in a quantized form $\hat{q}_n^k$.

\begin{figure}[t]
\centering
\includegraphics[scale=0.7]{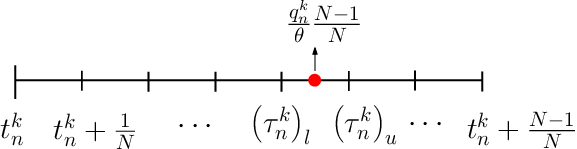} ~~
\includegraphics[scale=0.7]{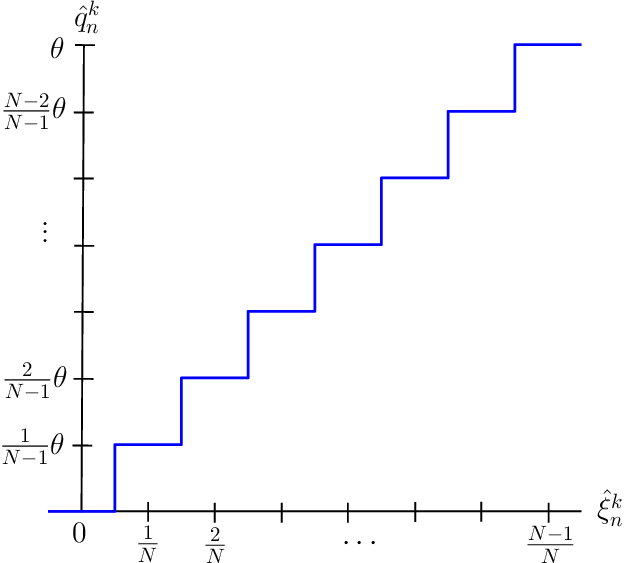}
\caption{The encoding and decoding mechanisms with finite time resolution $N$. In encoding $q_n^k$ into $\xi_n^k=\tau_n^k-t_n^k$, a pulse is transmitted at time either $\left(\tau_n^k\right)_l$ with probability $p$, given in \eqref{eq:quant_finite}, or $\left(\tau_n^k\right)_u$ with probability $1-p$. The decoder maps the estimate $\hat{\xi}_n^k$ to the corresponding quantization level.}
\label{fig:finite}
\end{figure}

Assume a system with $N$ available slots in a unit time interval. We select the discrete set of $\xi_n^k$ values as $\left\{ 0,\frac{1}{N},\ldots,\frac{N-1}{N} \right\}$ with $N$ elements.
Then, we accordingly determine $N$ quantization levels $\left\{ 0,\frac{\theta}{N-1},\ldots,\theta \right\}$ for $q_n^k$ by partitioning the range $[0,\theta)$ of $q_n^k$ into $N-1$ subintervals. When the overshoot falls into one of these subintervals, we quantize it into either the lower or upper end of the subinterval according to a randomization rule. Specifically, when $q_n^k\in[j\frac{\theta}{N-1},(j+1)\frac{\theta}{N-1}),~j=0,\ldots,N-2$, we quantize it into
\begin{align}
\label{eq:quant_finite}
    \hat{q}_n^k =& \left\{
    \ba{lll}
        j\frac{\theta}{N-1} & \text{with probability} & p=\frac{1-\exp\left(q_n^k-(j+1)\frac{\theta}{N-1}\right)}{1-\exp\left(-\frac{\theta}{N-1}\right)} \\
        (j+1)\frac{\theta}{N-1} & \text{with probability} & 1-p=\frac{\exp\left(q_n^k-j\frac{\theta}{N-1}\right)-1}{\exp\left(\frac{\theta}{N-1}\right)-1}
    \ea \right.,\\
    & \text{where} ~~~j=\left\lfloor \frac{q_n^k}{\theta/(N-1)} \right\rfloor, \nn
\end{align}
and $\lfloor\cdot\rfloor$ is the floor function. The reason why the randomization probability $p$ has this specific form is explained in Appendix. We then encode $\hat{q}_n^k$ into $\xi_n^k$, which is given by \be
\label{eq:delay_finite}
    \xi_n^k = \left\{
    \ba{lll}
        \frac{j}{N} & \text{with probability} & p \\
        \frac{j+1}{N} & \text{with probability} & 1-p
    \ea \right.,
\ee
where $p$ is given in \eqref{eq:quant_finite}. That is, we transmit a pulse for $b_n^k$ at time $\tau_n^k=t_n^k+\frac{j}{N}$ with probability $p$ or $\tau_n^k=t_n^k+\frac{j+1}{N}$ with probability $1-p$, as shown in Fig. \ref{fig:finite}. This time-encoding technique, not necessarily with the randomization rule, is called pulse position modulation (PPM).

At the other end of channel, the FC, upon receiving the pulse for $b_n^k$ at time $\tilde{\tau}_n^k$, estimates the transmission delay with $\hat{\xi}_n^k$. Note that $\hat{\xi}_n^k\in\bR$, unlike $\xi_n^k$. We decode $\hat{\xi}_n^k$ similarly to the encoding mechanism. Specifically, when $\hat{\xi}_n^k\in\left[\frac{2j-1}{2N}, \frac{2j+1}{2N}\right),~j=0,\ldots,N-1$, using the staircase function in Fig. \ref{fig:finite}, the FC recovers the quantized overshoot value $\hat{q}_n^k=j\frac{\theta}{N-1}$. Then, from \eqref{eq:change_enc}, it computes an approximate value $\hat{\lambda}_n^k$ of $\lambda_n^k$ using $b_n^k$ and $\hat{q}_n^k$. Say the $n$th message $\lambda_n^k$ from sensor $k$ is received by the FC as the $m$th message $\hat{\lambda}_m$ in the global order at time $\tilde{\tau}_m$. The FC then updates the approximate global LLR $\hat{L}_t$ as
\be
\label{eq:up_finite}
	\hat{L}_{\tilde{\tau}_m} = \hat{L}_{\tilde{\tau}_{m-1}} + \hat{\lambda}_m,
\ee
and keeps it constant until the arrival of the next message. After each update, the stopping and decision rules in \eqref{eq:FC_stop} and \eqref{eq:FC_dec}, respectively, are employed, as a result of which the sequential test either continues or stops and makes a decision between the hypotheses $\Hyp_0$ and $\Hyp_1$.

\subsection{Discussions}

Note that the FC uses \eqref{eq:up_finite} for all messages since we use the same parameters $\Delta$, $\theta$, and $N$ for all sensors. Hence, the FC does not need to identify the sensor from which a message originates.
To this end, we made the following assumptions.
\bi
\item[] \textbf{(A1)} There exists a global clock running in the wireless sensor network, hence the FC knows the potential sampling instants $t \in \bN$.
\item[] \textbf{(A2)} Overshoots are bounded by a constant $\theta \triangleq \max_{k,n} q_n^k$.
\item[] \textbf{(A3)} The FC reliably recovers the transmitted bit $b_n^k$, and estimates the transmission delay $\xi_n^k$ well enough so that the estimation error $|\xi_n^k-\hat{\xi}_n^k|<\frac{1}{2N}$.
\ei

We need (A3) for successful decoding of $\hat{\xi}_n^k$ (cf. Fig. \ref{fig:finite}).
In fact, (A3) implies that the estimation error $|\chi_n^k-\hat{\chi}_n^k|<\frac{1}{2N}$ where $\chi_n^k$ is the channel delay, random in general. The FC, upon receiving the pulse for $b_n^k$ at time $\tilde{\tau}_n^k$, in fact, estimates the channel delay (i.e., time-of-flight of the pulse) with $\hat{\chi}_n^k$, which then gives the estimates for transmission time $\hat{\tau}_n^k=\tilde{\tau}_n^k-\hat{\chi}_n^k$ and transmission delay $\hat{\xi}_n^k=\hat{\tau}_n^k-\lfloor\hat{\tau}_n^k\rfloor$. The constraint $|\chi_n^k-\hat{\chi}_n^k|<\frac{1}{2N}$ becomes stringent as the bound gets tighter for large $N$, which corresponds to a high time resolution case. In such a case, sensors need to transmit very short  pulses, requiring ultra-wideband (UWB) communications. Fortunately, in UWB, channel delay estimation can be performed very accurately \cite{UWB06}, since at least some of the frequencies have a line-of-sight trajectory, and UWB is robust against multipath fading due to the extremely short duration of pulses. Other advantages of UWB include compliance with strict energy constraints and robustness against eavesdropping \cite{UWB06}.

The UWB technology enables high data rate over short ranges. However, in our detector, in general, we do not need high accuracy in overshoot quantization for moderate error probability values (cf. Fig. \ref{fig:overs_comp}), hence we can trade high date rate offered by UWB for extension in range or for accuracy in recovering $b_n^k$ through coding. We can also increase the power level within the available budget for the same goals. We do not specify a modulation technique to transmit the sign bit $b_n^k$, for which PPM, pulse amplitude modulation (PAM) and binary phase shift keying (BPSK) are the most popular alternatives in a UWB system. In short, considering a UWB system, (A3) is a reasonable assumption. As an alternative to UWB, we can consider optical communications, in which PPM is commonly used, for the proposed detector.

\subsection{Asymptotic Analysis}

In the following theorem, we analyze the asymptotic average detection delay performance of the proposed detector with finite resolution. Below, $\cT$ and $\hat{\cT}$ are the stopping times of the optimal SPRT and the proposed detector, respectively, both of which satisfy the false alarm probability $\alpha$ and mis-detection probability $\beta$.

\begin{thm}
\label{thm:finite}
	The proposed sequential distributed detector with time resolution $N$ and randomization probability $p$ as in \eqref{eq:quant_finite} achieves the order-1 asymptotic optimality, i.e.,
    \be
        \frac{\Exp_i[\hat{\cT}]}{\Exp_i[\cT]}=1+o(1), ~i=0,1,~ \text{as}~ \alpha,\beta\to0,\nn
    \ee
    if the average message rate $R\to0$ [cf. \eqref{eq:delta}] at a rate slower than $\frac{1}{|\log \gamma_i|}$, i.e., $R|\log \gamma_i|\to\infty$, where $\gamma_0=\beta$, $\gamma_1=\alpha$, and $o(1)$ denotes a vanishing term as $\alpha,\beta\to0$. Moreover, it achieves the order-2 asymptotic optimality, i.e.,
    \be
        \Exp_i[\hat{\cT}]-\Exp_i[\cT]=O(1),~i=0,1,~ \text{as}~ \alpha,\beta\to0,\nn
    \ee
    with a constant $R$ if the time resolution $N\to\infty$ at least as fast as $|\log \gamma_i|$, i.e., $\frac{N}{|\log \gamma_i|}$ is bounded away from zero.
\end{thm}

The proof is given in Appendix. From \eqref{eq:delta}, we can rephrase the condition for the order-1 asymptotic optimality as $\Delta\to\infty$ at a rate slower than $|\log \gamma_i|$ since $\tanh\left(\frac{\Delta}{2}\right)\in(0,1)$ for $\Delta>0$ and $\sum_{k=1}^K |\Exp_i[L_1^k]|\not=0$ in a nontrivial case. In particular, with a finite time resolution, the quantization errors $|\lambda_m-\hat{\lambda}_m|$ accumulate without bound in time [cf. \eqref{eq:up_finite}], i.e., $|L_t-\hat{L}_t| \to\infty$ as $t\to\infty$. Sending less and less messages (i.e., larger and larger $\Delta$) asymptotically helps us control this error accumulation. On the other hand, too low message rate (i.e., too large $\Delta$) causes lack of information at the FC, which also degrades the asymptotic average detection delay performance. Hence, there is a trade-off in selecting the sampling threshold $\Delta$ when the time resolution is finite. The first part of Theorem \ref{thm:finite} gives the guidelines for setting $\Delta$ to achieve the order-1 asymptotic optimality.

The second part of Theorem \ref{thm:finite} says that an ever increasing time resolution, i.e., bandwidth, is needed for the order-2 asymptotic optimality. The minimum convergence rate is also specified. Fortunately, a UWB system can easily meet such a requirement for practical purposes where a limited range of target error probabilities are of interest.
\ignore{Note that order-2 implies order-1, which is the most commonly used type in the literature, but also the weakest type. The average detection delay of an order-1 scheme can diverge from that of the optimal scheme, whereas under order-2 optimality the average detection delay remains parallel to that of the optimal scheme.}

\subsection{Simulation Results}

\begin{figure}[t]
\centering
\includegraphics[scale=0.7]{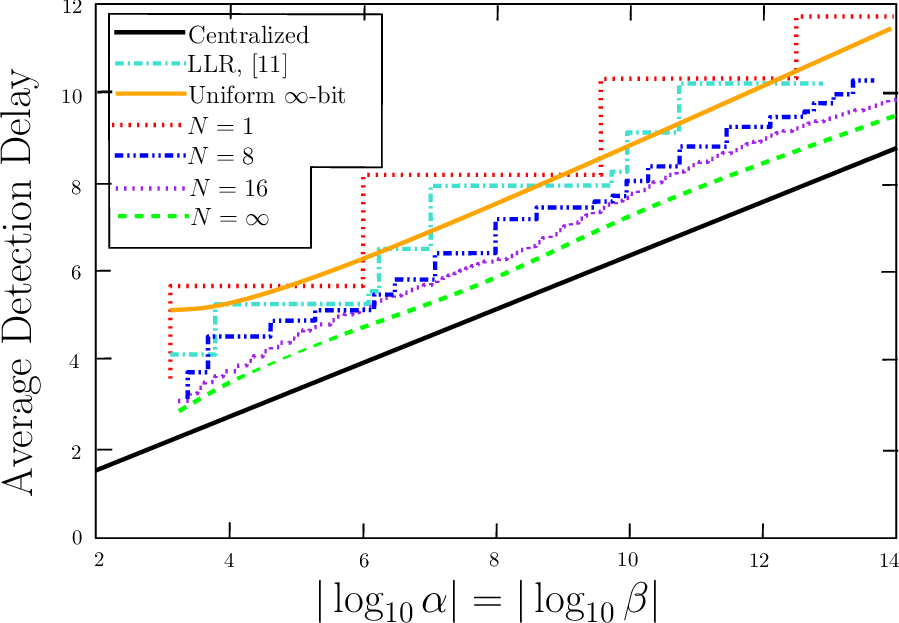}
\caption{Comparisons of the average detection delay performances of seven detectors.}
\label{fig:overs_comp}
\end{figure}

In Fig. \ref{fig:overs_comp}, the asymptotic average detection delay performances of seven detectors are compared, including
the scheme proposed in Section \ref{sec:time_enc} with $N=1,8,16,\infty$ time resolution, the scheme in \cite{Fellouris11}, the optimum centralized scheme, i.e., SPRT, and the scheme based on traditional uniform-in-time sampling with infinite number of bits. Note that the $N=1$ case simply ignores the overshoot problem.

In this example, we consider a system with two sensors, i.e., $K=2$, and an FC. Under $\Hyp_1$ the sensors observe i.i.d. $y_t^k \sim \cN(\sqrt[4]{10},1)$, whereas under $\Hyp_0$ they observe i.i.d. $y_t^k \sim \cN(0,1)$. Sensors use \cite[Eq. (16)]{Yilmaz12} to compute their local LLRs, which they report to the FC. The threshold $\Delta$ in the level-triggered sampling procedure is such that the average sampling time interval is the same as the sampling period $T=4$ of the uniform sampling procedure. All results are obtained by averaging $10^4$ trials and importance sampling is employed to compute probabilities of rare events. In Fig. \ref{fig:overs_comp}, the average detection delay performances are plotted under $\Hyp_1$.

The proposed detector with infinite time resolution, as expected, achieves the best performance, and its performance curve is parallel to that of the optimum centralized scheme, achieving the order-2 asymptotic optimality. Furthermore, the proposed scheme with a reasonably low time resolution ($N=16$) performs close to the infinite-resolution scheme, and achieves order-2 asymptotic optimality for practical purposes.
One remarkable result is that the $N=1$ scheme, which ignores the overshoot problem, outperforms the conventional uniform-sampling-based scheme with infinite number of bits at its achievable error rates, which was observed also in \cite{Yilmaz12}. The detectors that use only a discrete set of values for LLR update [cf. \eqref{eq:up_finite}], i.e., $N=1,8,16$ and the scheme in [11], can achieve only a discrete set of error rates, hence they have staircase performance curves. A detailed explanation of this phenomenon can be found in \cite[Fig. 1]{Yilmaz12}. Such a staircase curve tends to be linear as the number of available values for LLR update increases (compare $N=1$ and $N=16$).

\section{Target Detection in Wireless Sensor Networks}

In this section, we deal with target detection in wireless sensor networks as an application for the sequential distributed detector proposed in the previous section. In this application, we consider sensors as receivers in a radar network, in which transmitters might be sensors or some larger devices. Thus, in the remainder of the paper, the term sensor refers to a receiver in a radar network. Such sensors collaborate through a fusion center (FC) under stringent energy constraints to reach a global detection decision. The use of wireless sensor networks in radar applications has been considered in the literature under the name of radar sensor networks, e.g., \cite{Dutta06,Liang11}, which is an intuitive name for such systems. Different from the existing works on radar sensor networks, in this work, sensors are not necessarily monostatic radars, which have a transmitter and a receiver on them. We deal with a more general multistatic radar system, in which transmitters and receivers are not necessarily collocated. In that sense, the radar system considered in this paper is a special case of the broadly defined multi-input multi-output (MIMO) radar concept, which has been extensively studied in the literature, e.g., \cite{MIMOradar09} -- \cite{Moustakides12}.

In MIMO radar, multiple transmitters and multiple receivers are employed to make inference about a target, and different transmitters emit different waveforms to achieve waveform diversity \cite{MIMOradar09}.
The transmitters and the receivers of a MIMO radar system can be spatially distributed over a wide area to take advantage of the spatial (angular) properties of an extended target with many scatterers \cite{Fishler06,Haimovich08}. MIMO radars with collocated antennas, i.e., transmitters and receivers, in which a target is modeled as a point, have also been extensively studied in the literature, e.g., \cite{Li07} and references therein. In this paper, we consider the former case of MIMO radars (with widely separated antennas) since we are interested in the distributed detection problem. The concept of MIMO radar with widely separated antennas, more or less, coincides with those of multistatic radar, multisite radar and netted radar, e.g., \cite{Chernyak98}. The common goal is to mitigate the fluctuations in the amplitude and the phase of the signals received from a target by placing the antennas with wide spatial separation so that the fluctuations observed at different receivers become uncorrelated. In fact, the conditions for uncorrelated fluctuations, which can be found in \cite[Sec. II-A]{Fishler06}, depend not only on the distances between the antennas, but also the distances of the antennas to the target, dimensions of the target and the carrier wavelength.

In MIMO radar with well separated antennas, the local information at different receivers are fused, i.e., jointly processed,  at some level by the fusion center (FC). The level of information fusion ranges from a brief summary, e.g., a few bits, of observations per receiver (\emph{decentralized}) to the complete set of observations (\emph{centralized}). For radar applications both decentralized detection \cite{Longo96,Goodman07,Sammartino09}, and centralized detection\cite{Fishler06,Haimovich08,Tajer10} have been investigated. In the decentralized detection approaches for radar applications, e.g., \cite{Longo96,Goodman07,Sammartino09}, each receiver makes its own local decision, which is then forwarded to the FC using a single bit. The FC fuses the local decisions to reach a global decision (\emph{decision fusion}). On the other hand, \emph{data fusion}, in which each receiver sends a form of its local observations usually after some local processing, is a compromise between decision fusion and centralized detection. In general, data fusion techniques (and centralized detection as the ideal case) offer better performance than decision fusion techniques, but they require higher communication bandwidth and energy between the receivers and the FC. Fortunately, the decentralized detection techniques based on level-triggered sampling, e.g., the one proposed in Section \ref{sec:time_enc}, possess the best features of decision fusion and data fusion. In particular, the detector given in Section \ref{sec:time_enc} has a very low energy requirement, similar to the decision fusion methods, by transmitting only a single pulse per sample, and provides high performance, similar to the data fusion methods, since it combines local test statistics instead of local decisions. Finally, most of the existing works on MIMO radar consider fixed-sample-size detection, e.g., \cite{MIMOradar09,Fishler06,Haimovich08,Li07,Longo96,Goodman07,Sammartino09,Tajer10,Moustakides12}, whereas here we focus on sequential detection, which is more powerful in terms of promptly detecting the target.

\subsection{Problem Formulation}

We consider a wireless sensor network, i.e., a MIMO radar with widely separated antennas, consisting of $M$ transmitters and $K$ receivers. The problem of interest is to detect an extended target with $Q$ independent scatterers, whose coordinates are $X_q \in \bR^3,~q=1,\ldots,Q$ with a center of gravity at $X_0 \in \bR^3$. We similarly denote the coordinates of the transmitters and receivers as $X^t_m \in \bR^3,~m=1,\ldots,M$ and $X^r_{k} \in \bR^3,~k=1,\ldots,K$, respectively. Each transmitter $m$ emits a narrow-band waveform with the carrier frequency $f_c$ and the baseband equivalent signal $\sqrt{\frac{E}{M}}~s_m(t)$, where $E$ is the total transmitted energy by all transmitters.

If there is no target, i.e., under the null hypothesis $\Hyp_0$, each receiver $k$ observes the additive white complex Gaussian noise $w_{k}(t) \sim \cN_c(0,\sigma_{k}^2)$. On the other hand, if a target is present, i.e., under the alternative hypothesis $\Hyp_1$, each receiver $k$, in addition to $w_{k}(t)$, observes the superposition of all waveforms emitted by the $M$ transmitters and reflected by the $Q$ scatterers. Hence, we have the following binary hypothesis testing problem
\begin{align}
\label{eq:hyp_test}
\begin{split}
	\Hyp_0 &:  ~y_{k}(t)=w_{k}(t), \\
	\Hyp_1 &:  ~y_{k}(t)=\sum_{m=1}^M \sum_{q=1}^Q y_{mk}^q(t) + w_{k}(t),~~ k=1,\ldots,K.
\end{split}
\end{align}
The signal $y_{mk}^q(t)$ is emitted from transmitter $m$, reflected by scatterer $q$ and given by
\be
\label{eq:obs_one}
    y_{mk}^q(t) = \sqrt{\frac{E}{M}} h_{mk}^q (d_{mk}^q)^{-\eta} s_m(t-D_{mk}^q),
\ee
where $h_{mk}^q$, $d_{mk}^q$, $D_{mk}^q$ are the channel coefficient, the distance traveled by the signal, and the time delay experienced by the signal, respectively, between transmitter $m$ and receiver $k$ through scatterer $q$, and $\eta$ is the path-loss exponent. In \eqref{eq:obs_one}, different from \cite{Fishler06,Haimovich08}, we included the path-loss effect, as in \cite{Tajer10}.
The channel coefficient $h_{mk}^q$ consists of the reflectivity factor $\zeta_q$, which is a complex random variable, of scatterer $q$ and the phase shift incurred due to the time delay $D_{mk}^q$ and is given by \cite{Fishler06}
\be
\label{eq:ch_coef}
   h_{mk}^q = \zeta_q \exp(-j 2\pi f_c D_{mk}^q).
\ee
\ignore{The reflectivity factors $\{\zeta_q\}$ are modeled as zero-mean, i.i.d., complex random variables with variance $\Exp\big[|\zeta_q|^2\big]=1/Q$ since the scatterers are independent and isotropic. Note that the average radar cross section (RCS) is written as $\sum_{q=1}^Q \Exp\big[|\zeta_q|^2\big]=1$, which is independent of the number of scatterers. }The distance $d_{mk}^q$ between transmitter $m$ and receiver $k$ through scatterer $q$ and the time delay $D_{mk}^q$ are given by
\be
\label{eq:dist_del}
	d_{mk}^q = \|X^t_m-X_q\|_2+\|X^r_{k}-X_q\|_2 ~~~\text{and}~~~ D_{mk}^q=d_{mk}^q/c,
\ee
respectively, where $c$ is the speed of light.

We assume that the distances from sensors to the target are significantly larger than the dimensions of the target, i.e., $\max_{p,q} \|X_q-X_p\|_2 \ll \min_{m,k,q} \|d_{mk}^q\|$, hence $d_{mk}^q=d_{mk}=\|X^t_m-X_0\|_2+\|X^r_{k}-X_0\|_2,\forall q$ and accordingly $D_{mk}^q=D_{mk}=d_{mk}/c$ as in \cite{Tajer10,Moustakides12}.
Then, from \eqref{eq:hyp_test} and \eqref{eq:obs_one} under $\Hyp_1$ each receiver $k$ observes
\be
\label{eq:obs_all}
	\Hyp_1 : ~y_{k}(t) = \sqrt{\frac{E}{M}} ~\sum_{m=1}^M h_{mk} ~d_{mk}^{-\eta} ~s_m(t-D_{mk}) + w_{k}(t),
\ee
where $h_{mk} \triangleq \sum_{q=1}^Q h_{mk}^q$. We assume that the coordinates $\{X^t_m\}$, $\{X^r_{k}\}$, $X_0$ and the wavelength $\lambda_c$ satisfy the sufficient condition similar to that in \cite{Fishler06} so that the channel coefficients $\{h_{mk}\}$ are uncorrelated. The sufficient condition in \cite{Fishler06} which was provided for the two-dimensional coordinates can be straightforwardly extended to the three-dimensional coordinates.

\subsection{Sequential Distributed Target Detection}

Each receiver $k$ samples its observed signal $y_{k}(t)$ at rate $1/T_s$ and obtains the discrete-time signal $y^{k}_t$. The discrete-time version of \eqref{eq:hyp_test} is then
\begin{align}
\label{eq:obs_disc}
\begin{split}
    \Hyp_0 &:  ~y^{k}_t = w^{k}_t \\
    \Hyp_1 &:  ~y^{k}_t = \sqrt{\frac{E}{M}} ~\sum_{m=1}^M h_{mk} ~d_{mk}^{-\eta} ~s^m_{t,D_{mk}} + w^{k}_t,
\end{split}
\end{align}
where $w^{k}_t \triangleq w_{k}(tT_s)$, and $s^m_{t,D} \triangleq s_m(tT_s-D)$. \ignore{We assume that the sampling rate is high enough so that the discrete-time signals $\{s^m_{t,\tau}\}_m$ remain orthogonal for any delay $\tau$ of interest.}

To perform sequential target detection for the wireless sensor network under consideration we employ the sequential distributed detector based on level-triggered sampling given in Section \ref{sec:time_enc}. Specifically, each receiver $k$ computes the LLR
\be
\label{eq:LLR_loc}
    L^{k}_t=\log\frac{f_1^{k}\big(\{y_{\tau}^{k}\}_{\tau=1}^t\big)}{f_0^{k}\big(\{y_{\tau}^{k}\}_{\tau=1}^t\big)},
\ee
samples it using level-triggered sampling [cf. \eqref{eq:samp_time} -- \eqref{eq:loc_dec}], sends a pulse per sample to the FC by encoding the overshoot in time [cf. \eqref{eq:delay_finite}]. The FC performs the SPRT-like test based on the received information from the sensors [cf. \eqref{eq:FC_stop}, \eqref{eq:FC_dec}, \eqref{eq:up_finite}]. In \eqref{eq:LLR_loc}, $f_0^{k}(\cdot)$ is the joint pdf of $\{y_{\tau}^{k}\}_{\tau}$ under $\Hyp_0$, which is $\cN_c(0,\sigma_{k}^2 \vec{I})$ from \eqref{eq:obs_disc}, where $\vec{I}$ is the $t\times t$ identity matrix. On the other hand, the joint pdf $f_1^{k}(\cdot)$ under $\Hyp_1$ depends on the distribution of the channel coefficients $\{h_{mk}\}_m$. In what follows we discuss how to compute $L^{k}_t$ for the four Swerling fluctuating target models \cite{Skolnik01}.

\subsubsection{Swerling case 1}

This model assumes that the targets consist of many independent scatterers of comparable echo areas, i.e., no scatterer is larger than the others. Accordingly, the reflectivity factors $\{\zeta_q\}$ are modeled as zero-mean i.i.d. complex random variables with variance $\Exp[|\zeta_q|^2]=1/Q$, \cite{Fishler06,Haimovich08,Tajer10}. Then, due to the central limit theorem the channel coefficient $h_{mk}=\sum_{q=1}^Q h_{mk}^q$ is distributed as $\cN_c(0,1)$ from \eqref{eq:ch_coef}. Moreover, $\{h_{mk}\}$ are independent as they are uncorrelated. In this model, $\{h_{mk}\}$ remain constant during the entire scan, but are independent over different scans.

The distances $\{d_{mk}\}$ and the time delays $\{D_{mk}\}$ are known with the coordinates $\{X^t_m\}$, $\{X^r_{k}\}$ and $X_0$ given [cf. \eqref{eq:dist_del}]. In this case, we have a composite hypothesis testing problem with the unknown parameters $\{h_{mk}\}$. There are two main approaches treating unknown parameters in SPRT. The first one, called the weighted SPRT (WSPRT), integrates out $\{h_{mk}\}_{m}$ over the joint pdf $f_1^{k}(\{y^{k}_{\tau}\}_{\tau},\{h_{mk}\}_{m})$ to obtain the likelihood $f_1^{k}(\{y^{k}_{\tau}\}_{\tau})$. Alternatively, the second approach, called the generalized sequential likelihood ratio test (GSLRT), replaces $f_1^{k}(\{y^{k}_{\tau}\}_{\tau})$ by the maximum of $f_1^{k}(\{y^{k}_{\tau}\}_{\tau},\{h_{mk}\}_{m})$ over $\{h_{mk}\}_{m}$.

\vspace{5mm}
\underline{{\it WSPRT:}}
Defining the vectors $\vh_{k} \triangleq [h_{1k},\ldots,h_{Mk}]^T$ and $\vy^{k}_t \triangleq [y_1^{k},\ldots,y_t^{k}]^T$ we write
\begin{align}
    f_1^{k}(\vy^{k}_t) &= \int_{\vh_{k}} f_1^{k}(\vy^{k}_t|\vh_{k}) f_{\vh_{k}}(\vh_{k}) ~\text{d}\vh_{k}, \label{eq:pdf_y} \\
    \text{where}~~ f_1^{k}(\vy^{k}_t|\vh_{k}) &= (\pi\sigma_{k}^2)^{-t} \exp\Bigg(\underbrace{-\frac{1}{\sigma_{k}^2} \sum_{\tau=1}^t \Big|y_{\tau}^{k}-\sqrt{\frac{E}{M}}\sum_{m=1}^M h_{mk} ~d_{mk}^{-\eta} ~s^m_{t,D_{mk}}\Big|^2}_{J(\vh_{k})} \Bigg), \label{eq:like1} \\
    \text{and}~~ f_{\vh_{k}}(\vh_{k}) &= \pi^{-M} \exp(-\|\vh_{k}\|^2), \label{eq:pdf_h}
\end{align}
where $\|\cdot\|$ is the Euclidean norm.
In \eqref{eq:like1}, we used \eqref{eq:obs_disc} and the fact that $\{w^{k}_{\tau}\}_{\tau}$ are i.i.d. and distributed as $\cN_c(0,\sigma_{k}^2)$. Similarly, to write \eqref{eq:pdf_h} we used the fact that $\{h_{mk}\}_m$ are i.i.d. and distributed as $\cN_c(0,1)$. Expanding $J$ in \eqref{eq:like1} we have
\begin{align}
    J &= -\frac{1}{\sigma_{k}^2}\sum_{\tau=1}^t \Big|y_{\tau}^{k}-\sqrt{\frac{E}{M}}\sum_{m=1}^M h_{mk} ~d_{mk}^{-\eta} ~s^m_{\tau,D_{mk}}\Big|^2 \nn\\
    &= -\frac{\|\vy^{k}_t\|^2}{\sigma_{k}^2} + \sum_{m=1}^M \Big[h_{mk}^*V_{mk}^t + h_{mk}(V_{mk}^t)^* - U_{mk}^t|h_{mk}|^2 - \sum_{n=1,n\not=m}^M Z_{mnk}^t h_{mk}h_{nk}^* \Big], \label{eq:cross}\\
    \begin{split}
    \text{with}~~ V_{mk}^t &\triangleq \sqrt{\frac{E}{M}} \frac{d_{mk}^{-\eta}}{\sigma_{k}^2} \sum_{\tau=1}^t y_{\tau}^{k}(s^m_{\tau,D_{mk}})^*,
    ~~ U_{mk}^t \triangleq \frac{E}{M} \frac{d_{mk}^{-2\eta}}{\sigma_{k}^2} \sum_{\tau=1}^t |s^m_{\tau,D_{mk}}|^2,\\
    \text{and}~~ Z_{mnk}^t &\triangleq \frac{E}{M} \frac{d_{mk}^{-\eta}d_{nk}^{-\eta}}{\sigma_{k}^2} \sum_{\tau=1}^t s^m_{\tau,D_{mk}}(s^n_{\tau,D_{nk}})^*, \label{eq:abc}
    \end{split}
\end{align}
where $^*$ denotes complex conjugate.

Substituting \eqref{eq:cross} in \eqref{eq:like1} and then using \eqref{eq:like1} and \eqref{eq:pdf_h} we can obtain $f_1^{k}(\vy^{k}_t)$ by computing the integral in \eqref{eq:pdf_y}. However, computing this $M$-dimensional integral is not feasible in general due to the cross-correlation term $Z_{mnk}^t$. In this case, we follow the common practice and assume that the transmitted waveforms are orthonormal \cite{Fishler06,Haimovich08,Tajer10}, i.e.,
\be
\label{eq:orth_assump}
	\int_{S} s_m(t) s_n^*(t-D)~\text{d}t=\delta(m-n),~m,n=1,\ldots,M,
\ee
for all time delays $D$ of interest, where $S$ is the duration of the waveforms, and $\delta(\cdot)$ is the Dirac's delta function. We further assume, as in \cite{Haimovich08,Tajer10}, that the sampling rate $1/T_s$ is high enough so that the discrete-time signals $\{s^m_{t,D_{mk}}\}_m$ remain orthogonal, i.e.,
\be
\label{eq:orth_assump_disc}
    \sum_{\tau=1}^{S_d} s^m_{\tau,D_{mk}}(s^n_{\tau,D_{nk}})^* = 0,~m,n=1,\ldots,M,~m\not=n,
\ee
and for all time delays $D_{mk}$ and $D_{nk}$ of interest, where $S_d=S/T_s$ is an integer for simplicity. Then, $Z_{mnk}^t$ disappears for $t=S_d,2S_d,\ldots$. In fixed-sample-size schemes, such as the ones in \cite{Fishler06,Haimovich08,Tajer10}, the LLR is computed once, whereas in the sequential scheme it is computed at every time instant. To reduce the computational complexity we propose that the receivers compute their local LLRs only at time instants $pS_d,~p\in\bN^+$, i.e., the global clock in the system is downscaled by $S_d$. Note that now the sampling times $\{t_n^k\}$ [cf. \eqref{eq:samp_time}] are given by
\be
\label{eq:samp_time_1}
	t_n^k = \min\{ pS_d\in\bN: |L_{pS_d}^k-L_{t_{n-1}^k}^k| \geq \Delta  \},~L_0^k=0,~p\in\bN^+,
\ee
where, instead of \eqref{eq:delta}, we should use the following equation
\be
\label{eq:delta_1}
    \Delta \tanh\left(\frac{\Delta}{2}\right)=\frac{1}{RS_d}\sum_{k=1}^K |\Exp_i[L_{S_d}^{k}]|
\ee
to compute $\Delta$ for an average message rate of $R$ messages per $T_s$ seconds. The FC computes the stopping time [cf. \eqref{eq:FC_stop}] as follows
\be
\label{eq:FC_stop_1}
    \hat{\cT} = \min\{ pS_d\in\bN: \hat{L}_{pS_d} \not\in (-\hat{B},\hat{A})\}.
\ee

Then, to obtain $L_{pS_d}^k$ we first write $f_1^k(\vy_{pS_d}^k)$ using \eqref{eq:pdf_y}--\eqref{eq:abc} as
\begin{align}
    f_1^{k}(\vy^{k}_{pS_d}) &= \frac{\exp\left(-\frac{\|\vy_{pS_d}^k\|^2}{\sigma_k^2}\right)}{\pi^{pS_d+M}\sigma_k^{2pS_d}} \int_{\vh_{k}} \exp\left( -\sum_{m=1}^M (U_{mk}^{pS_d}+1)|h_{mk}^2| - h_{mk}^*V_{mk}^{pS_d} - h_{mk}(V_{mk}^{pS_d})^* \right)  \text{d}\vh_{k} \nn\\
    &= \frac{\exp\left( \sum_{m=1}^M \frac{|V_{mk}^{pS_d}|^2}{U_{mk}^{pS_d}+1}-\frac{\|\vy_{pS_d}^k\|^2}{\sigma_k^2} \right)}{\pi^{pS_d}\sigma_k^{2pS_d} \prod_{m=1}^M (U_{mk}^{pS_d}+1)} \nn\\
    &~~~~~~~~~~~~~ \underbrace{\int_{\vh_{k}} \frac{\prod_{m=1}^M (U_{mk}^{pS_d}+1)}{\pi^M} \exp\left( -\sum_{m=1}^M (U_{mk}^{pS_d}+1) \Bigg| h_{mk}-\frac{V_{mk}^{pS_d}}{U_{mk}^{pS_d}+1} \Bigg|^2 \right) \text{d}\vh_{k}}_{=1} \nn\\
    &= \frac{\exp\left( \sum_{m=1}^M \frac{\left|\sum_{\tau=1}^{pS_d} y_{\tau}^{k}(s^m_{\tau,D_{mk}})^*\right|^2}{\sigma_{k}^2\left(\frac{p}{T_s}+\frac{M}{E} \frac{\sigma_k^2}{d_{mk}^{-2\eta}}\right)} - \frac{\|\vy_{pS_d}^k\|^2}{\sigma_k^2} \right)} {\pi^{pS_d}\sigma_k^{2pS_d} \prod_{m=1}^M \left(\frac{E}{M} \frac{p}{T_s} \frac{d_{mk}^{-2\eta}}{\sigma_k^2} +1\right)}, \label{eq:pdf_y_1}
\end{align}
where we used the definitions of $V_{mk}^t$ and $U_{mk}^t$, given in \eqref{eq:abc}, and $\sum_{\tau=1}^{pS_d} |s^m_{\tau,D_{mk}}|^2 = \frac{p}{T_s}$ from the Riemann sum for the integral in \eqref{eq:orth_assump}. Since $f_0^{k}(\vy^{k}_{pS_d})=\frac{1}{\pi^{pS_d}\sigma_k^{2pS_d}}\exp\left( -\frac{\|\vy_{pS_d}^k\|^2}{\sigma_k^2} \right)$, we write the LLR $L_{pS_d}^k = \log \frac{f_1^{k}(\vy^{k}_{pS_d})}{f_0^{k}(\vy^{k}_{pS_d})}$ as
\be
\label{eq:LLR_s1}
	L_{pS_d}^k = \sum_{m=1}^M \left[ \frac{\left|\sum_{\tau=1}^{pS_d} y_{\tau}^{k}(s^m_{\tau,D_{mk}})^*\right|^2}{\sigma_{k}^2\left(\frac{p}{T_s}+\frac{M}{E} \frac{\sigma_k^2}{d_{mk}^{-2\eta}}\right)} - \log \left(\frac{E}{M} \frac{p}{T_s} \frac{d_{mk}^{-2\eta}}{\sigma_k^2} +1\right) \right],
\ee
where the inner product $\sum_{\tau=1}^{pS_d} y_{\tau}^{k}(s^m_{\tau,D_{mk}})^*$ can be regarded as the output of a matched filter, which can be used to compute an estimate of $h_{mk}$ as a result of the orthogonality condition in \eqref{eq:orth_assump_disc}.

To summarize, under the Swerling case 1 target model in WSPRT, each receiver $k$ computes $L_{pS_d}^k$ as in \eqref{eq:LLR_s1} at each time $pS_d, ~p\in\bN^+$, samples it at times $\{t_n^k\}$, given in \eqref{eq:samp_time_1}, and sends bits $\{b_n^k\}$ [cf. \eqref{eq:loc_dec}] at times $\{\tau_n^k\}$ [cf. \eqref{eq:delay_finite}] to the FC. Then, the FC upon receiving a bit $b_n$ at time $\tau_n$ updates its approximate LLR $\hat{L}_t$ as in \eqref{eq:up_finite}, terminates the procedure according to \eqref{eq:FC_stop_1}, and makes its final decision using \eqref{eq:FC_dec}.
Note that the resulting scheme is a ``slow" version of the scheme introduced in Section \ref{sec:time_enc}, hence Theorem \ref{thm:finite} still holds.

\vspace{5mm}
\underline{{\it GSLRT:}}
In GSLRT, $L_t^k$ is given by
\be
\label{eq:LLR_g}
	L_t^k = \log \frac{\max_{\vh_{k}} f_1^{k}(\vy^{k}_t|\vh_{k}) f_{\vh_{k}}(\vh_{k})}{f_0^{k}(\vy^{k}_t)},
\ee
where the maximum \emph{a posteriori} (MAP) estimator $\hat{\vh}_k$ maximizes the numerator. Using \eqref{eq:like1}--\eqref{eq:abc} we compute $\hat{\vh}_k$ as
\begin{align}
\label{eq:MAP}
	\hat{\vh}_k &= \arg\max_{\vh_{k}} f_1^{k}(\vy^{k}_t|\vh_{k}) f_{\vh_{k}}(\vh_{k}) \nn\\
	&= \arg\min_{\vh_k} \vh_k^H \mG_t^k \vh_k -\vh_k^H \va_t^k - (\va_t^k)^H \vh_k \nn\\
	&= (\mG_t^k)^{-1} \va_t^k,
\end{align}
where
\be
\label{eq:Ga}
	\mG_t^k \triangleq \MB{c c c c} U_{1k}^t+1 & Z_{12k}^t & \cdots & Z_{1Mk}^t \\
							Z_{21k}^t & U_{2k}^t+1 & \cdots & Z_{2Mk}^t \\
							\vdots & \vdots & \ddots & \vdots \\
							Z_{M1k}^t & Z_{M2k}^t & \cdots & U_{Mk}^t+1 \ME
	, ~~~ \va_t^k \triangleq \MB{c} V_{1k}^t \\ V_{2k}^t \\ \vdots \\ V_{Mk}^t  \ME,
\ee
and $(\cdot)^H$ denotes the Hermitian transpose of a vector or matrix. Note from the definition of $Z_{mnk}^t$ in \eqref{eq:abc} that $\mG_t^k$ is a Hermitian matrix, i.e., $\mG_t^k=(\mG_t^k)^H$. Then, from \eqref{eq:like1}--\eqref{eq:cross}, \eqref{eq:MAP} and \eqref{eq:Ga} we write $L_t^k$ as
\begin{align}
	L_t^k &= \log \frac{(\pi\sigma_{k}^2)^{-t} \pi^{-M} \exp\left( -\frac{\|\vy^{k}_t\|^2}{\sigma_{k}^2} - \hat{\vh}_k^H \mG_t^k \hat{\vh}_k + \hat{\vh}_k^H \va_t^k + (\va_t^k)^H \hat{\vh}_k \right)} {(\pi\sigma_{k}^2)^{-t} \exp\left( -\frac{\|\vy^{k}_t\|^2}{\sigma_{k}^2} \right)} \nn\\
	&= (\va_t^k)^H (\mG_t^k)^{-1} \va_t^k - M\log\pi. \label{eq:GSLRT}
\end{align}
Note that we did not need the orthogonality assumption in \eqref{eq:orth_assump_disc} to write the above equation, which is used to compute $L_t^k$ at each time $t$. In this general form, without the orthogonality assumption, each receiver $k$ needs to compute $(\mG_t^k)^{-1}$, the inverse of an $M\times M$ matrix, at each time $t$. Specifically, with \eqref{eq:orth_assump_disc} the off-diagonal terms, $\{Z_{mnk}^t\}$, vanish at times $pS_d,~p\in\bN^+$, hence at these time instants \eqref{eq:GSLRT} takes the following simple form
\be
\label{eq;GSLRT_orth}
	L_{pS_d}^k = \sum_{m=1}^M \frac{\left|\sum_{t=1}^{pS_d} y_{\tau}^{k}(s^m_{\tau,D_{mk}})^*\right|^2}{\sigma_{k}^2\left(\frac{p}{T_s}+\frac{M}{E} \frac{\sigma_k^2}{d_{mk}^{-2\eta}}\right)} - M\log\pi,
\ee
which is similar to \eqref{eq:LLR_s1} written for WSPRT. Hence, in GSLRT we have several options. As the first option, each sensor $k$ can compute $L_t^k$ as in \eqref{eq:GSLRT} by inverting $\mG_t^k$ [cf. \eqref{eq:Ga}] at each time $t$ and follow the scheme in Section \ref{sec:time_enc}. Alternatively, assuming \eqref{eq:orth_assump_disc} it may compute $L_t^k$ only at $t=pS_d,~p\in\bN^+$ as in \eqref{eq;GSLRT_orth}, avoiding the computationally expensive matrix inversion operation, and follow the ``slow" scheme introduced for WSPRT. Another option is a compromise between the first two; that is, each sensor $k$ computes $L_t^k$ using \eqref{eq:GSLRT} at each time $t$ with the assumption in \eqref{eq:orth_assump_disc}, where \eqref{eq;GSLRT_orth} becomes a special case of \eqref{eq:GSLRT}. Note that Theorem \ref{thm:finite} does not hold for GSLRT since the test statistic $L_t^k$ has different forms in \eqref{eq:LLR_loc} and \eqref{eq:LLR_g}.

\vspace{5mm}
\subsubsection{Swerling case 2}

This model differs from the first one only in the fluctuation rate. Specifically, under this model $\{h_{mk}\}$ are again i.i.d. and distributed as $\cN_c(0,1)$, but they change much more quickly. Here we assume that $\{h_{mk}^t\}$ are independent for each time $t$.
Then, from \eqref{eq:obs_disc} under $\Hyp_1$, $y_t^k \sim \cN_c(0,\rho_k^2+\sigma_k^2)$ where $\rho_k^2 \triangleq \frac{E}{M} \sum_{m=1}^M d_{mk}^{-2\eta} |s_{t,D_{mk}}^m|^2$ and for simplicity $|s_{t,D_{mk}}^m|^2$ is the same for all $t$. Consequently, the LLR is given by
\begin{align}
\label{eq:rho}
	L_t^k &= \log \frac{[\pi(\rho_k^2+\sigma_{k}^2)]^{-t} \exp\left( -\frac{\|\vy^{k}_t\|^2}{\rho_k^2+\sigma_{k}^2} \right)} {(\pi\sigma_{k}^2)^{-t} \exp\left( -\frac{\|\vy^{k}_t\|^2}{\sigma_{k}^2} \right)} \nn\\
	&= \frac{\rho_k^2}{(\rho_k^2+\sigma_{k}^2)\sigma_{k}^2} \|\vy^{k}_t\|^2 + t \log \frac{\sigma_{k}^2}{\rho_k^2+\sigma_{k}^2} \nn\\
	&= L_{t-1}^k + \underbrace{\frac{\rho_k^2}{(\rho_k^2+\sigma_{k}^2)\sigma_{k}^2} |y_t^k|^2 + \log \frac{\sigma_{k}^2}{\rho_k^2+\sigma_{k}^2}}_{l_t^k} = \sum_{\tau=1}^t l_{\tau}^k.
\end{align}
Note that the orthogonality assumption is not required to write \eqref{eq:rho}. Furthermore, we are able to compute $L_t^k$ recursively. Under this target model, using \eqref{eq:rho} we follow the detector proposed in Section \ref{sec:time_enc}.

\vspace{5mm}
\subsubsection{Swerling case 3}

This model is proposed for targets consisting of a large scatterer together with a number of small scatterers, all of which are independent. Here we consider the Rician target model \cite[page 72]{Skolnik01}.
The reflectivity factors $\{\zeta_q\}$ of the small scatterers are modeled as zero-mean, i.i.d. complex random variables with variance $1/Q$, and that of the large scatterer has the same distribution with the same variance as the others, but a with a nonzero mean. Then, the channel coefficients $\{h_{mk}\}$ are i.i.d. and distributed as $\cN_c(\mu,1)$ due to the central limit theorem and from \eqref{eq:ch_coef}. In this model, $\{h_{mk}\}$ remain constant during the entire scan, as in the first model. Hence, similar to the first model we can treat this case either using WSPRT or GSLRT.

\vspace{5mm}
\underline{{\it WSPRT:}}
Assuming \eqref{eq:orth_assump_disc}, we write the likelihood $f_1^k(\vy_t^k)$ at times $pS_d,~p\in\bN^+$ as
\be
    f_1^k(\vy_{pS_d}^k) = \frac{\exp\left( \sum_{m=1}^M \frac{|V_{mk}^{pS_d}+\mu|^2}{U_{mk}^{pS_d}+1}-\frac{\|\vy_{pS_d}^k\|^2}{\sigma_k^2} - M|\mu|^2 \right)}{\pi^{pS_d}\sigma_k^{2pS_d} \prod_{m=1}^M (U_{mk}^{pS_d}+1)}
\ee
similar to \eqref{eq:pdf_y_1}. Hence, the LLR is given by
\be
\label{eq:LLR_3}
	L_{pS_d}^k = \log \frac{f_1^{k}(\vy^{k}_{pS_d})}{f_0^{k}(\vy^{k}_{pS_d})} = \sum_{m=1}^M \left[ \frac{|V_{mk}^{pS_d}+\mu|^2}{U_{mk}^{pS_d}+1} - |\mu|^2 - \log \left(U_{mk}^{pS_d}+1\right) \right],
\ee
where $V_{mk}^{pS_d}=\sqrt{\frac{E}{M}} \frac{d_{mk}^{-\eta}}{\sigma_{k}^2} \sum_{\tau=1}^t y_{\tau}^{k}(s^m_{\tau,D_{mk}})^*$ and $U_{mk}^{pS_d}=\frac{E}{M} \frac{p}{T_s} \frac{d_{mk}^{-2\eta}}{\sigma_k^2}$. Then, using \eqref{eq:LLR_3} the ``slow" scheme explained for the first model can be similarly obtained.

\vspace{5mm}
\underline{{\it GSLRT:}}
The MAP estimator $\hat{\vh}_k$ is written as
\begin{align}
\label{eq:MAP_3}
	\hat{\vh}_k &= \arg\min_{\vh_k} \vh_k^H \mG_t^k \vh_k -\vh_k^H \tilde{\va}_t^k - (\tilde{\va}_t^k)^H \vh_k \nn\\
	&= (\mG_t^k)^{-1} \tilde{\va}_t^k,
\end{align}
where $\tilde{\va}_t^k \triangleq \va_t^k+\mu$, $\mG_t^k$ and $\va_t^k$ are given in \eqref{eq:Ga}. Using $\hat{\vh}_k$ we write the LLR $L_t^k$ as
\be
\label{eq:GSLRT_3}
	L_t^k = (\tilde{\va}_t^k)^H (\mG_t^k)^{-1} \tilde{\va}_t^k - M(|\mu|^2+\log\pi),
\ee
where the orthogonality assumption is not needed as in \eqref{eq:GSLRT}. With the assumption in \eqref{eq:orth_assump_disc}, at times $pS_d,~p\in\bN^+$  \eqref{eq:GSLRT_3} is simplified to
\be
\label{eq;GSLRT_orth_3}
	L_{pS_d}^k = \sum_{m=1}^M \frac{|V_{mk}^{pS_d}+\mu|^2}{U_{mk}^{pS_d}+1} - M(|\mu|^2+\log\pi).
\ee
As in the first model with GSLRT we can either follow the scheme in Section \ref{sec:time_enc} using \eqref{eq:GSLRT_3} or
the ``slow" scheme using \eqref{eq;GSLRT_orth_3} or a compromise between them.

\vspace{5mm}
\subsubsection{Swerling case 4}

This target model is the same as the third model except for its faster fluctuation rate. Similar to the second model we assume that $\{h_{mk}^t\}$ are independent for each time $t$. That is, in this model $\{h_{mk}^t\}$ are i.i.d. with the distribution $\cN_c(\mu,1)$. Hence, we have
\begin{align}
\label{eq:rho_4}
	L_t^k &= \log \frac{[\pi(\rho_k^2+\sigma_{k}^2)]^{-t} \exp\left( -\frac{\|\vy^{k}_t-\vec{\tilde{\mu}}_t^k\|^2}{\rho_k^2+\sigma_{k}^2} \right)} {(\pi\sigma_{k}^2)^{-t} \exp\left( -\frac{\|\vy^{k}_t\|^2}{\sigma_{k}^2} \right)} \nn\\
	&= L_{t-1}^k + \underbrace{\frac{\frac{\rho_k^2}{\sigma_{k}^2} |y_t^k|^2 + (y_t^k)^*\tilde{\mu}_t^k + (\tilde{\mu}_t^k)^* y_t^k - |\tilde{\mu}_t^k|^2}{\rho_k^2+\sigma_{k}^2} + \log \frac{\sigma_{k}^2}{\rho_k^2+\sigma_{k}^2}}_{l_t^k} = \sum_{\tau=1}^t l_{\tau}^k,
\end{align}
where $\vec{\tilde{\mu}}_t^k=[\tilde{\mu}_1^k,\ldots,\tilde{\mu}_t^k]^T$, $\tilde{\mu}_t^k \triangleq \mu \sqrt{\frac{E}{M}} \sum_{m=1}^M d_{mk}^{-\eta} s^m_{t,D_{mk}}$, and $\rho_k^2$ is as defined in the second model. Using \eqref{eq:rho_4} we follow the scheme given in Section \ref{sec:time_enc}.

\subsection{Simulation Results}

In this section, we provide simulation results for the target models discussed in the previous subsection. The existing works, to the best of our knowledge, all consider fixed-sample-size tests to perform (centralized or decentralized) detection in MIMO radar. Since there is no direct way of comparison between sequential and fixed-sample-size tests, and the existing works on decentralized detection for MIMO radar consider only the decision fusion techniques, we compare our schemes with the sequential schemes that apply decision fusion techniques. Specifically, in the schemes that we compare with, sensors run individual SPRTs, whose decisions are then fused at the FC using the majority rule, which can be seen as the extensions of the fixed-sample-size tests in the literature dealing with decentralized detection for MIMO radar.

All results are obtained by averaging $10^4$ trials. Each transmitter emits the waveform
$$s_m(t)=(1/\sqrt{S}) \exp(j2\pi m t /S) [U(t)-U(t-S)],$$
where $U(t)$ is the unit step function, with the waveform duration $S=2 \times 10^{-7}$ sec., i.e.,
with the carrier frequency $f_c=5$ MHz. The path loss coefficient is $\eta=2$ as in free space, and the sampling period at the receivers is $T_s=S/2$, i.e., $S_d=2$. The level-triggered sampling threshold $\Delta$ is determined to ensure an average message rate of $R=K/4$ messages per $T_s$ seconds [cf. \eqref{eq:delta} and \eqref{eq:delta_1}], where $K$ denotes the number of receivers. The mean of the channel coefficients in the last two target models is set as $\mu=(1+j)/3$. In all models the noise variance $\sigma_k^2=1,\forall k$ is used for all receivers. We assume, as in \cite{Tajer10}, that the transmitters, the receivers and the target are located at $X_m^t=(m,0,0)$, $X_k^r=(0,k,0)$ and $X_0=(20,15,0)$, respectively, where all the distances are in km. We present four sets of simulations to compare the average detection delay performances of the centralized scheme, the proposed decentralized scheme with resolution $N=16$ and the decision fusion scheme that uses the majority rule to combine local decisions at the FC. In particular, the latter declares $\Hyp_1$ if the number of local decisions that are in favor of $\Hyp_1$ is larger than $K/2$. In the subsequent figures, we plot the average detection delay performances under $\Hyp_1$, which is of primary concern in radar applications. Similar performances are observed under $\Hyp_0$.

\subsubsection{\textbf{\underline{Fixed SNR, $K$ and $M$, varying $\alpha,\beta$}}}

\begin{figure}[t]
\centering
\includegraphics[scale=0.5]{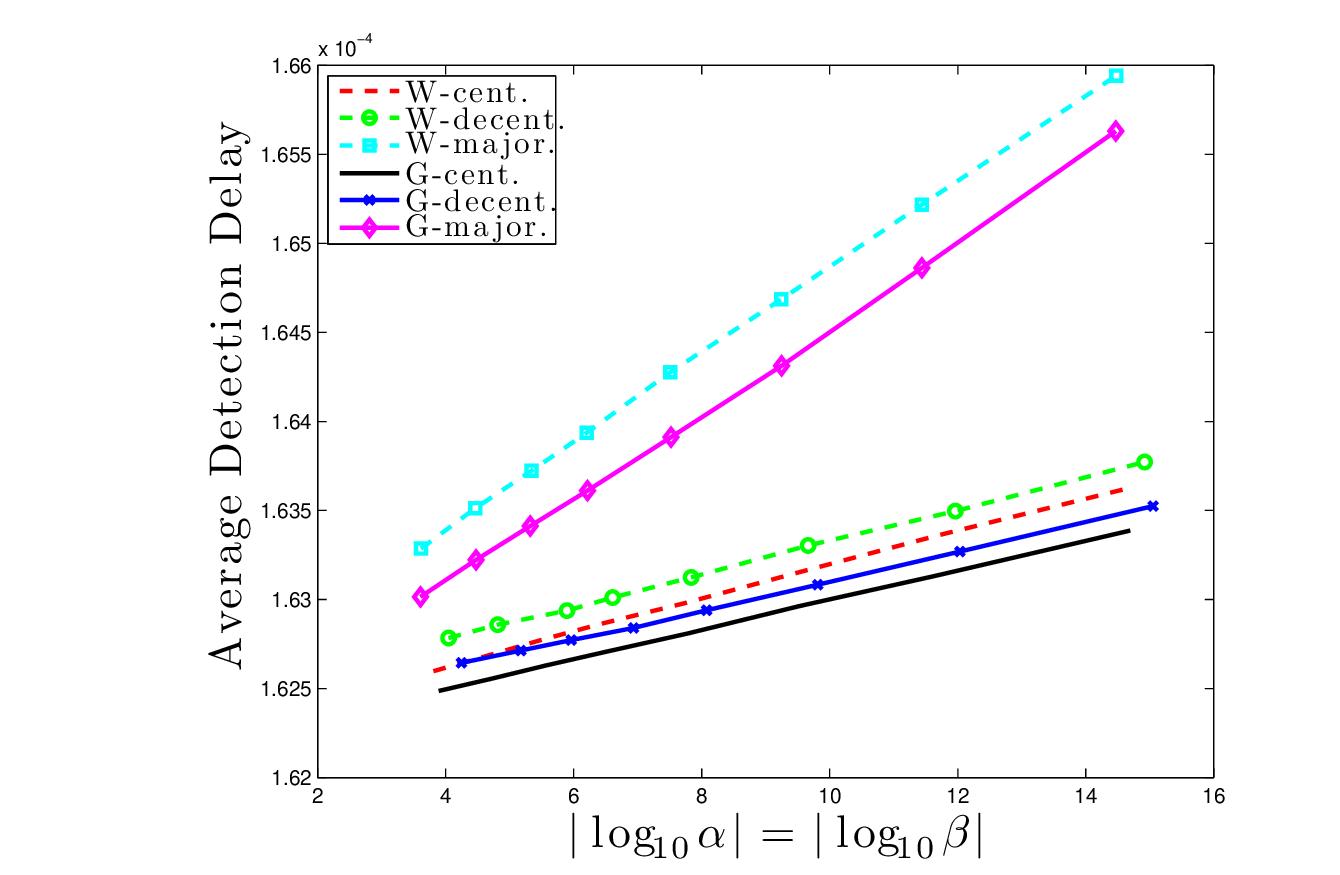}
\caption{Average detection delay vs. error probabilities for the optimum centralized scheme, the decentralized scheme based on level-triggered sampling and the decision fusion scheme with majority rule that are proposed for the Swerling case 1 target model and use WSPRT and GSLRT.}
\label{fig:S1_err}
\end{figure}

In the first set of simulations, we analyze the average detection delay performances under different false alarm and mis-detection probabilities $\alpha$ and $\beta$, respectively. We assume a MIMO radar with two transmitters and two receivers, i.e., $K=M=2$, for which we set SNR$=E/\sigma_k^2=2$ ($3$ dB) and vary the error probabilities $\alpha$ and $\beta$ together between $10^{-1}$ and $10^{-15}$.

Fig. \ref{fig:S1_err} illustrates the asymptotic performances for the Swerling case 1 target model using both WSPRT and GSLRT. It is seen that the proposed decentralized schemes that use WSPRT and GSLRT both exhibit order-2 asymptotic optimality for practical purposes, as their performance curves are parallel to those of the centralized schemes. In fact, the performance curves of the decentralized schemes are not perfectly linear as shown in Fig. \ref{fig:overs_comp}. Nevertheless, with resolution $N=16$, this effect is not remarkable (cf. Fig. \ref{fig:overs_comp}), hence ignored here.
There is a significant difference between the asymptotic performances of the proposed decentralized schemes and decision fusion schemes, which are the conventional methods used for decentralized detection in MIMO radar.
We observe that under $\Hyp_1$ the GSLRT-based schemes perform better than the WSPRT-based schemes since the joint pdf $f_1^{k}(\vy^{k}_t,\vh_{k})$ under $\Hyp_1$ is maximized in GSLRT, whereas it is averaged in WSPRT. The opposite relationship holds under $\Hyp_0$. For the other three target models we observe similar results to those noted for Swerling case 1.

\begin{figure}[t]
\centering
\includegraphics[scale=0.5]{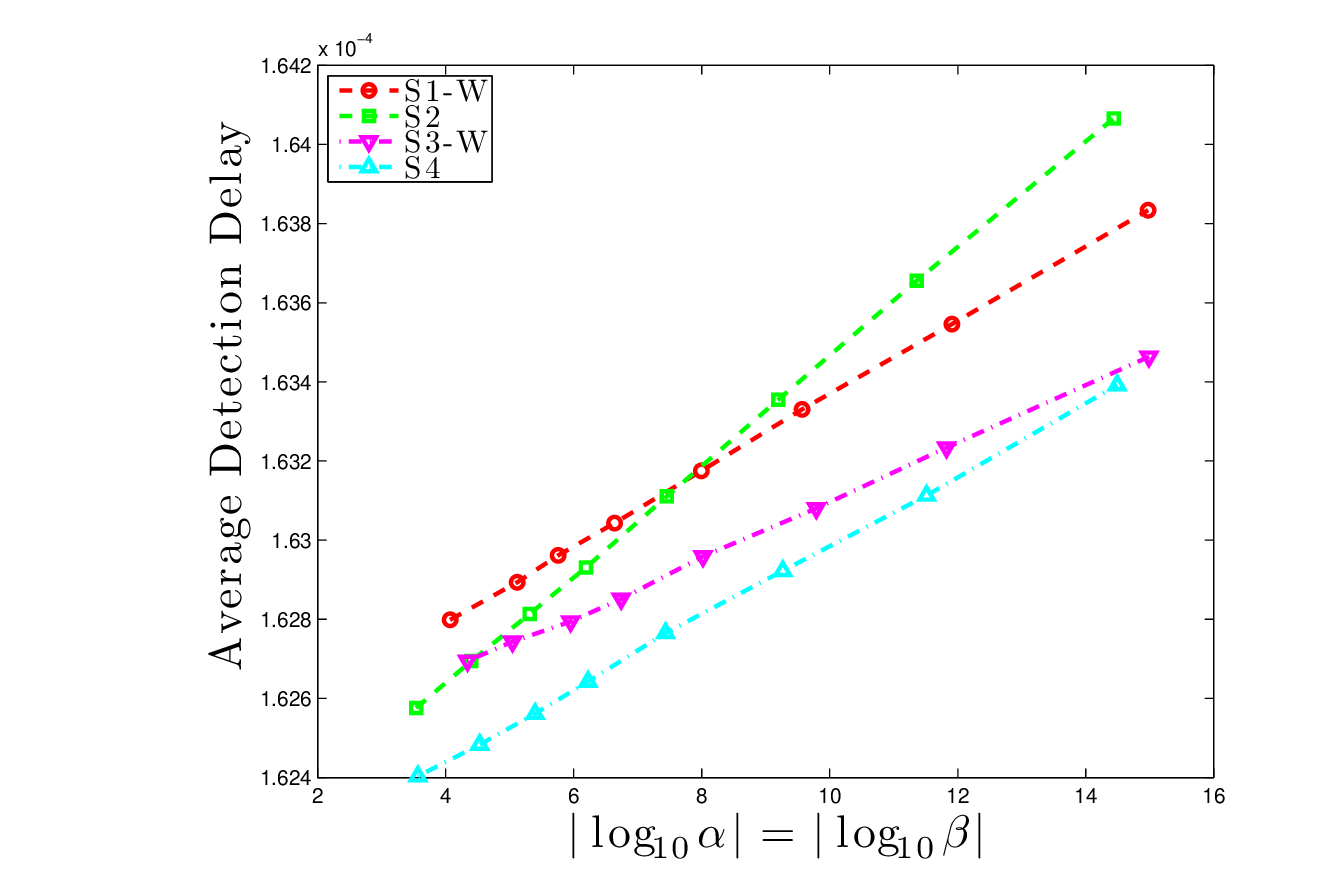}
\caption{Average detection delay vs. error probabilities for the decentralized schemes proposed for the four Swerling target models. }
\label{fig:Sall_err}
\end{figure}

In Fig. \ref{fig:Sall_err}, the asymptotic performances of the decentralized detectors proposed for the four target models are compared. The WSPRT-based schemes are shown for the first and the third target models since the marginal pdf $f_1^{k}(\vy^{k}_t)$ under $\Hyp_1$ is used in WSPRT as in the schemes proposed for the second and the fourth models. It is seen that the schemes proposed for the third and the fourth models benefit from signals with nonzero means as their performance curves lie below those of the schemes proposed for the first and the second models and also have smaller slopes. Moreover, as a result of fast fluctuations the performance curves of the schemes proposed for the second and the fourth models have larger slopes than those for the first and the third models. On the other hand, the curves that correspond to the fast fluctuating (second and fourth) models partially and completely lie below the curves that correspond to the slow fluctuating first and third models, respectively, for $\alpha,\beta \in [10^{-15},10^{-1}]$.

\subsubsection{\textbf{\underline{Fixed $\alpha,\beta$, $K$ and $M$, varying SNR}}}

We next consider the average detection delay performances under different SNR conditions with fixed $\alpha=\beta=10^{-6}$ and $K=M=2$.

\begin{figure}[t]
\centering
\includegraphics[scale=0.5]{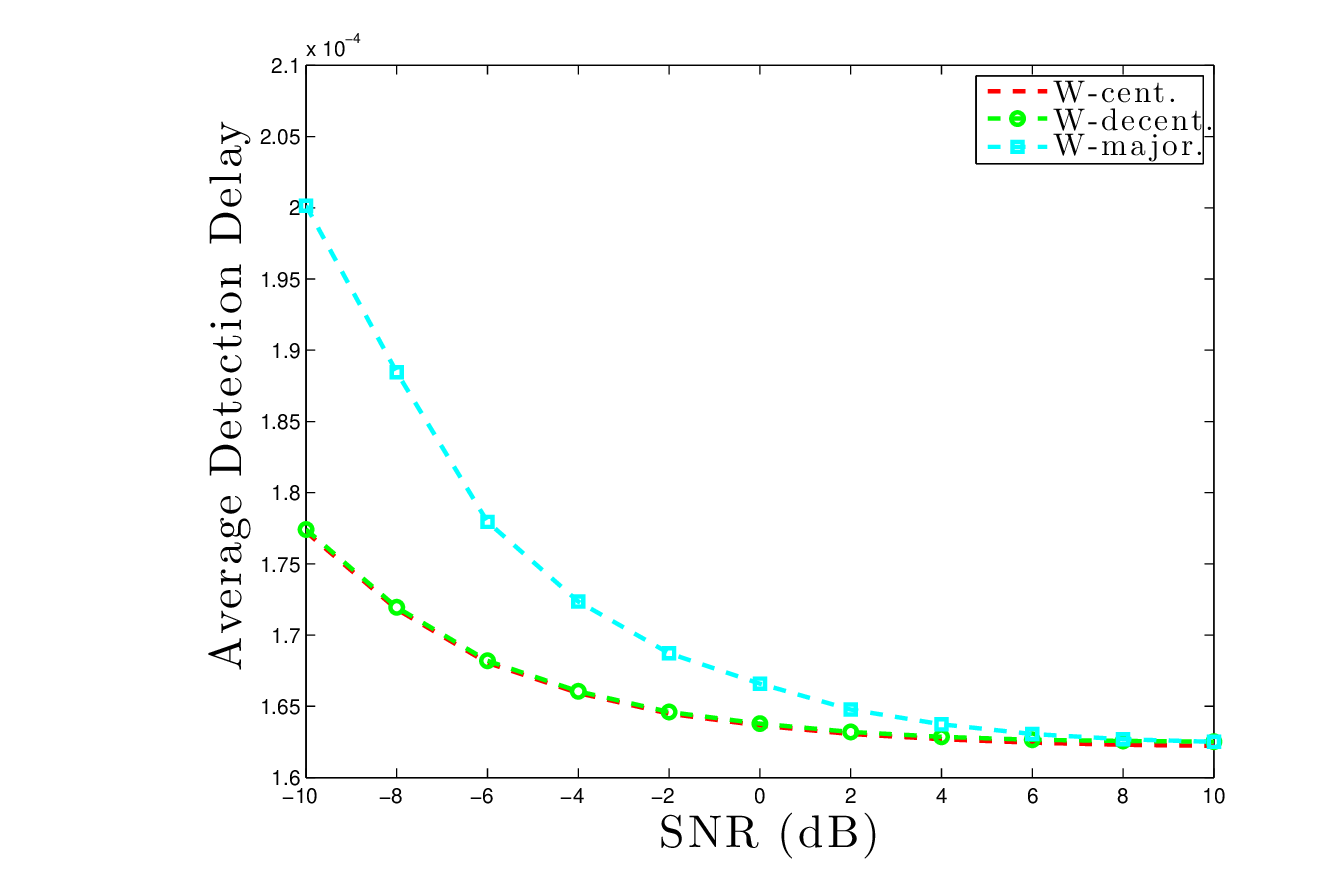}
\caption{Average detection delay vs. SNR for the optimum centralized scheme, the decentralized scheme based on level-triggered sampling and the decision fusion scheme with majority rule, proposed for the Swerling case 1 target model.}
\label{fig:S1_snr}
\end{figure}

It is shown in Fig. \ref{fig:S1_snr} that the WSPRT-based decentralized scheme proposed for the first target model achieves a very close performance to that of the corresponding centralized scheme for different SNR values. Whereas the performance of the decision fusion scheme is much worse than that of the centralized scheme especially for low SNR values. For high SNR values all schemes achieve similar performances. Similar observations hold for the GSLRT-based schemes proposed for the first target model and also for the other schemes proposed for the other three target models.

\begin{figure}[t]
\centering
\includegraphics[scale=0.5]{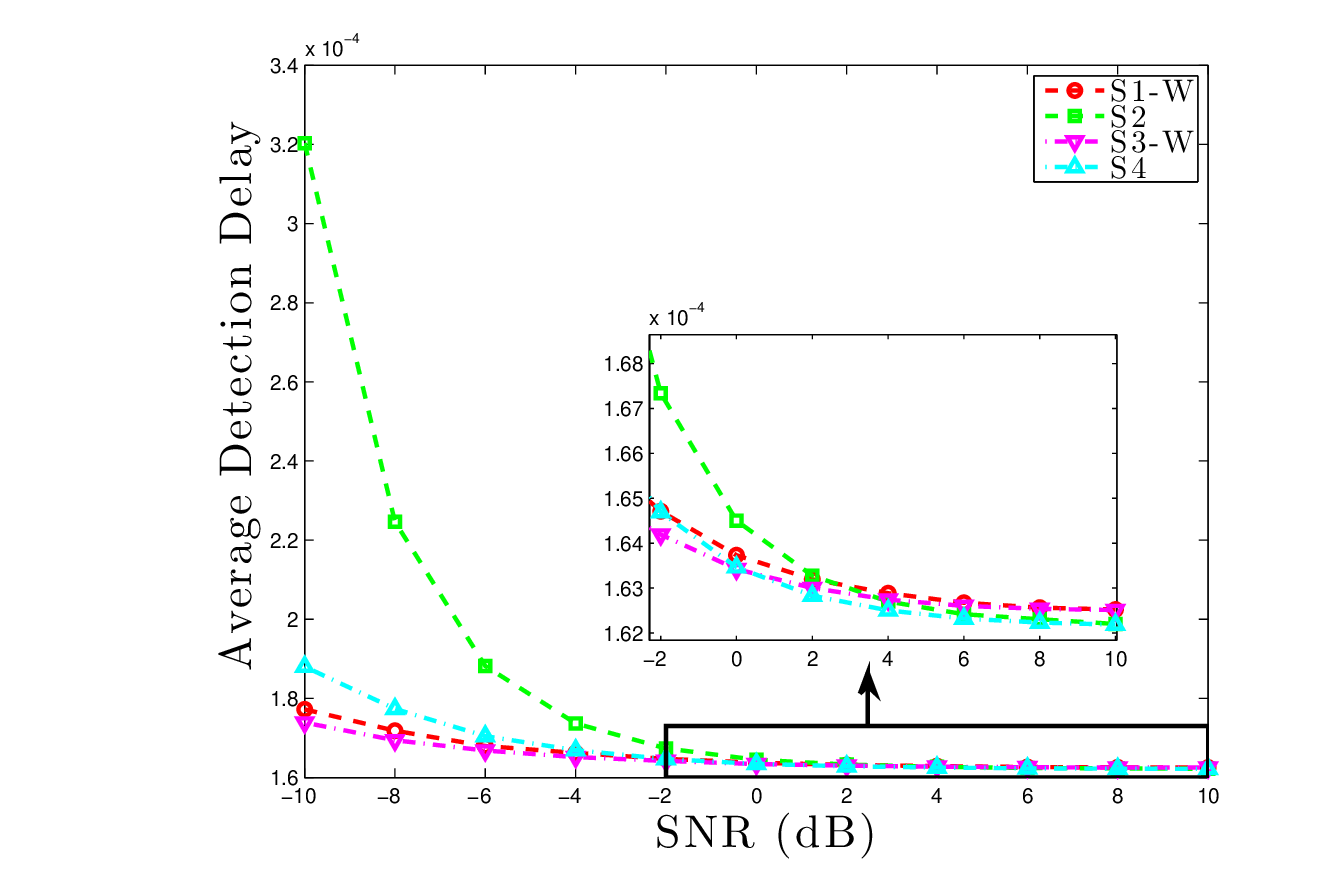}
\caption{Average detection delay vs. SNR$=E/\sigma_k^2$ for the decentralized schemes proposed for the four Swerling target models.}
\label{fig:Sall_snr}
\end{figure}

Fig. \ref{fig:Sall_snr} compares the average detection delay performances of the decentralized schemes proposed for the four target models under different SNR conditions. While the fast fluctuating (second and fourth) target models perform worse than the slow fluctuating ones for low SNR values, the opposite is true for high SNR values as shown in the zoomed version of the figure. This result is in accordance with Fig. 2.22 in \cite[page 67]{Skolnik01}. The models that represent targets with a large scatterer, i.e., third and fourth target models, have better performances than the ones without any large scatterer, i.e., first and second target models, respectively, for all SNR values.

\subsubsection{\textbf{\underline{Fixed $\alpha,\beta$, SNR and $M$, varying $K$}}}

Next we analyze the increasing receiver diversity case where $\alpha=\beta=10^{-6}$, SNR$=3$dB, $M=2$, and $K=2,\ldots,10$.

\begin{figure}[t]
\centering
\includegraphics[scale=0.5]{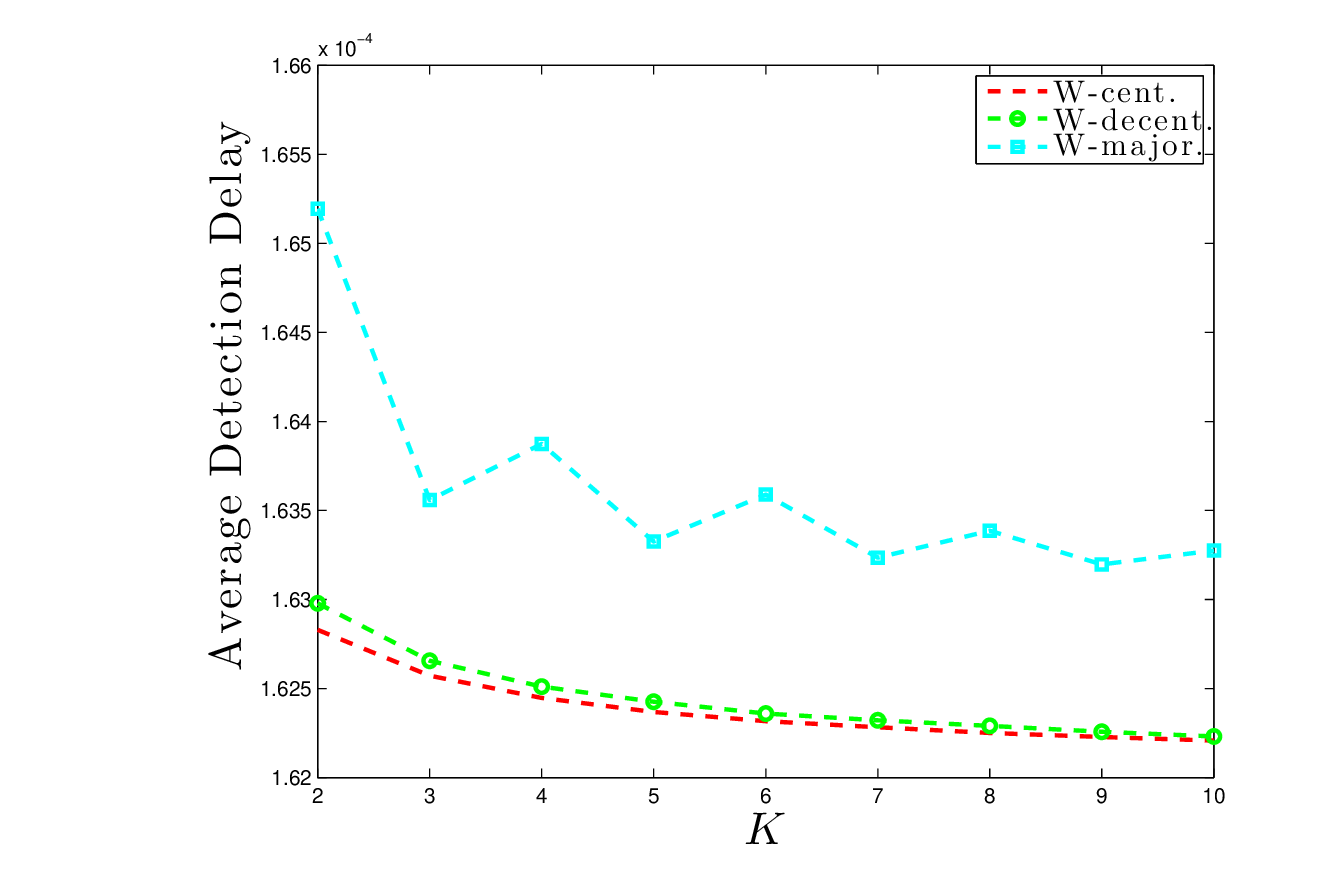}
\caption{Average detection delay vs. the number of receivers $K$ for the optimum centralized scheme, the decentralized scheme based on level-triggered sampling and the decision fusion scheme with majority rule, proposed for the Swerling case 1 target model.}
\label{fig:S1_K}
\end{figure}

The average detection delays of the centralized and the proposed decentralized schemes decrease with the increasing receiver diversity, as shown in Fig. \ref{fig:S1_K}. For the decision fusion scheme, the average detection delay monotonically decreases for odd $K$ or even $K$. But the delay corresponding to an even $K=2k$ is
larger than the preceding odd $K=2k-1$.
 This discrepancy is because in the majority rule for even $K$ values higher percentages of $K$ receivers must decide on $\Hyp_1$ to declare $\Hyp_1$ than those for the neighboring odd $K$ values. Same observations hold for the other target models.

\begin{figure}[t]
\centering
\includegraphics[scale=0.5]{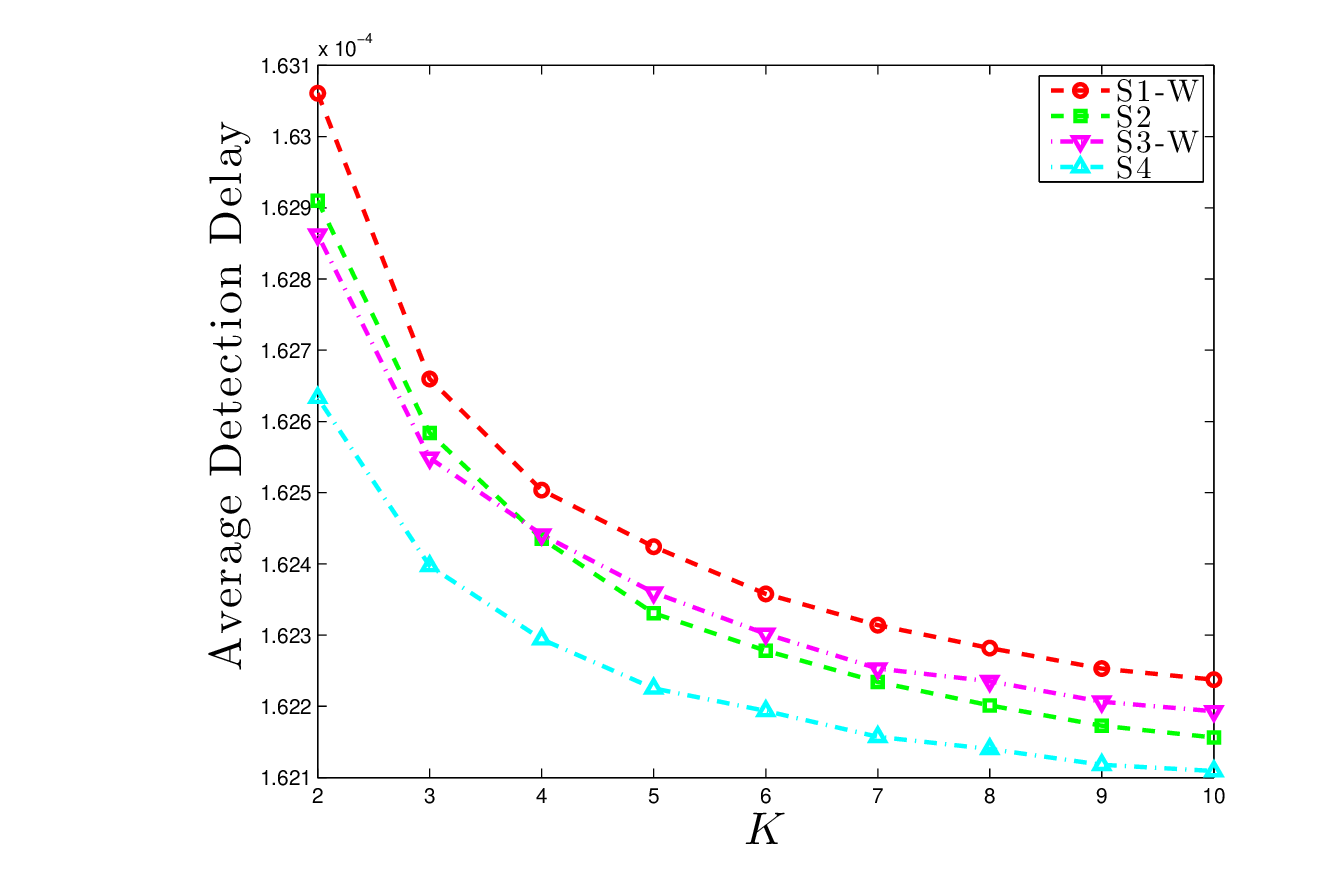}
\caption{Average detection delay vs. the number of receivers $K$ for the decentralized schemes proposed for the four Swerling target models.}
\label{fig:Sall_K}
\end{figure}

In Fig. \ref{fig:Sall_K} it is seen that the fast fluctuating (second and fourth) models enjoy the increasing receiver diversity more than the slow fluctuating (first and third) models. Moreover, nonzero means of the channel coefficients in the third and the fourth models improve the average detection delay performances.

\subsubsection{\textbf{\underline{Fixed $\alpha,\beta$, SNR and $K$, varying $M$}}}

In the final set of simulations, increasing transmitter diversity is considered. We set $\alpha=\beta=10^{-6}$, SNR$=3$dB, $K=2$, and $M$ ranges from $2$ to $40$.

\begin{figure}[t]
\centering
\includegraphics[scale=0.5]{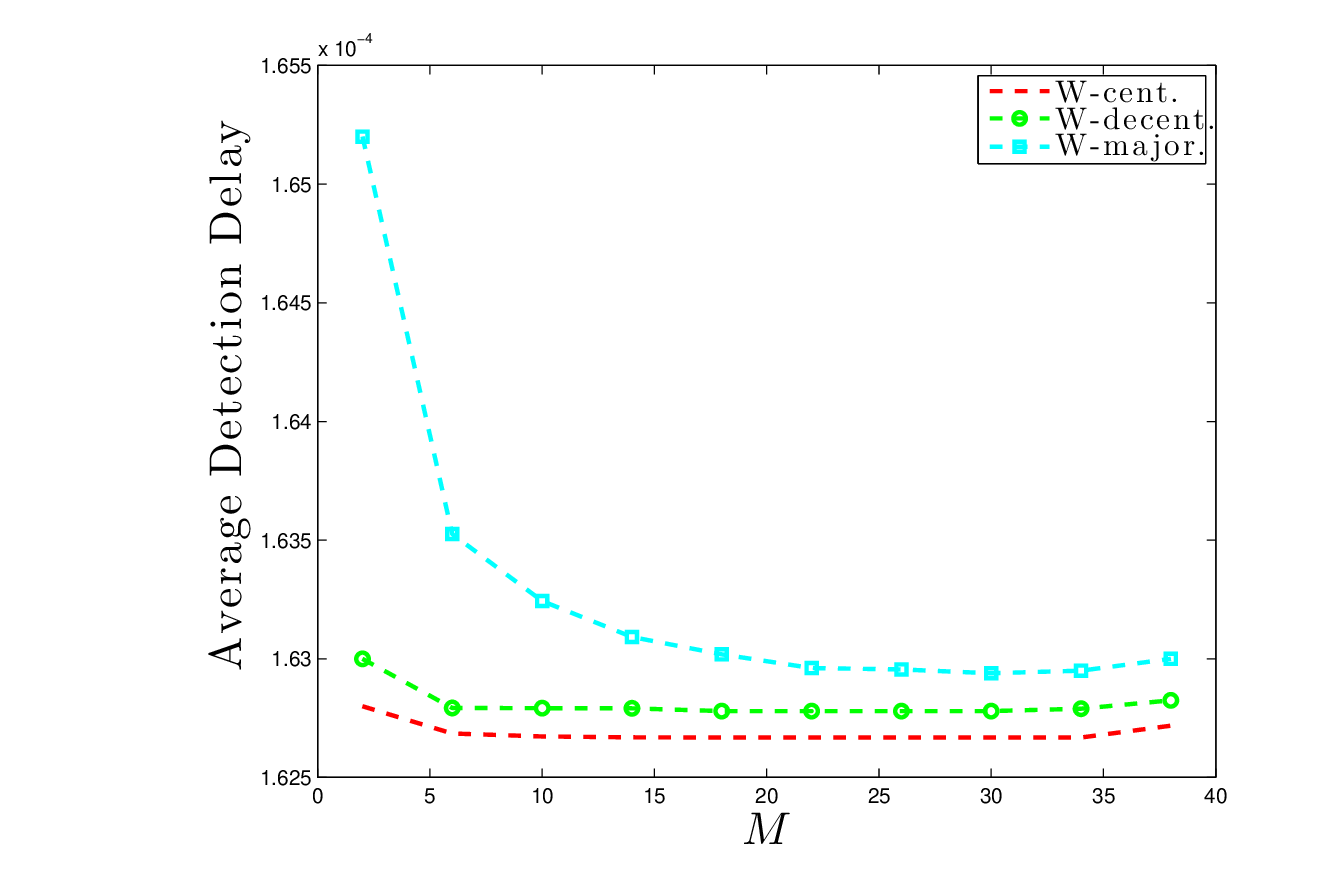}
\caption{Average detection delay vs. the number of transmitters $M$ for the optimum centralized scheme, the decentralized scheme based on level-triggered sampling and the decision fusion scheme with majority rule, proposed for the Swerling case 1 target model.}
\label{fig:S1_M}
\end{figure}

Note that the total power $E$ emitted from the transmitters remains the same for fixed SNR$=E/\sigma_k^2$ although the number of transmitters increases. Hence, the performances of the centralized scheme and the proposed decentralized scheme tend to stay constant as shown in Fig. \ref{fig:S1_M}. However, we observe in Fig. \ref{fig:S1_M} that the performance curves of those schemes change for small and large $M$ values due to the network topology. Since the target and the transmitters are located at $X_0=(20,15,0)$ and $X_m^t=(m,0,0)$, respectively, the newly added transmitters increase and decrease the average LLR when $M$ is small and large, respectively. This is because the distance between the target and the newly added transmitter, and thus the path loss decreases and increases when $M$ is small and large, respectively. On the other hand, the performance of the decision fusion scheme is affected more by changing $M$ since it is governed by the individual receiver characteristics. When the average distance between the target and the transmitters is small, e.g., $M=30$, the average path loss is small, the average LLR is large, and all schemes perform similarly. This result is similar to the high SNR case, hence expected. Same observations again hold for the other target models.

\begin{figure}[t]
\centering
\includegraphics[scale=0.5]{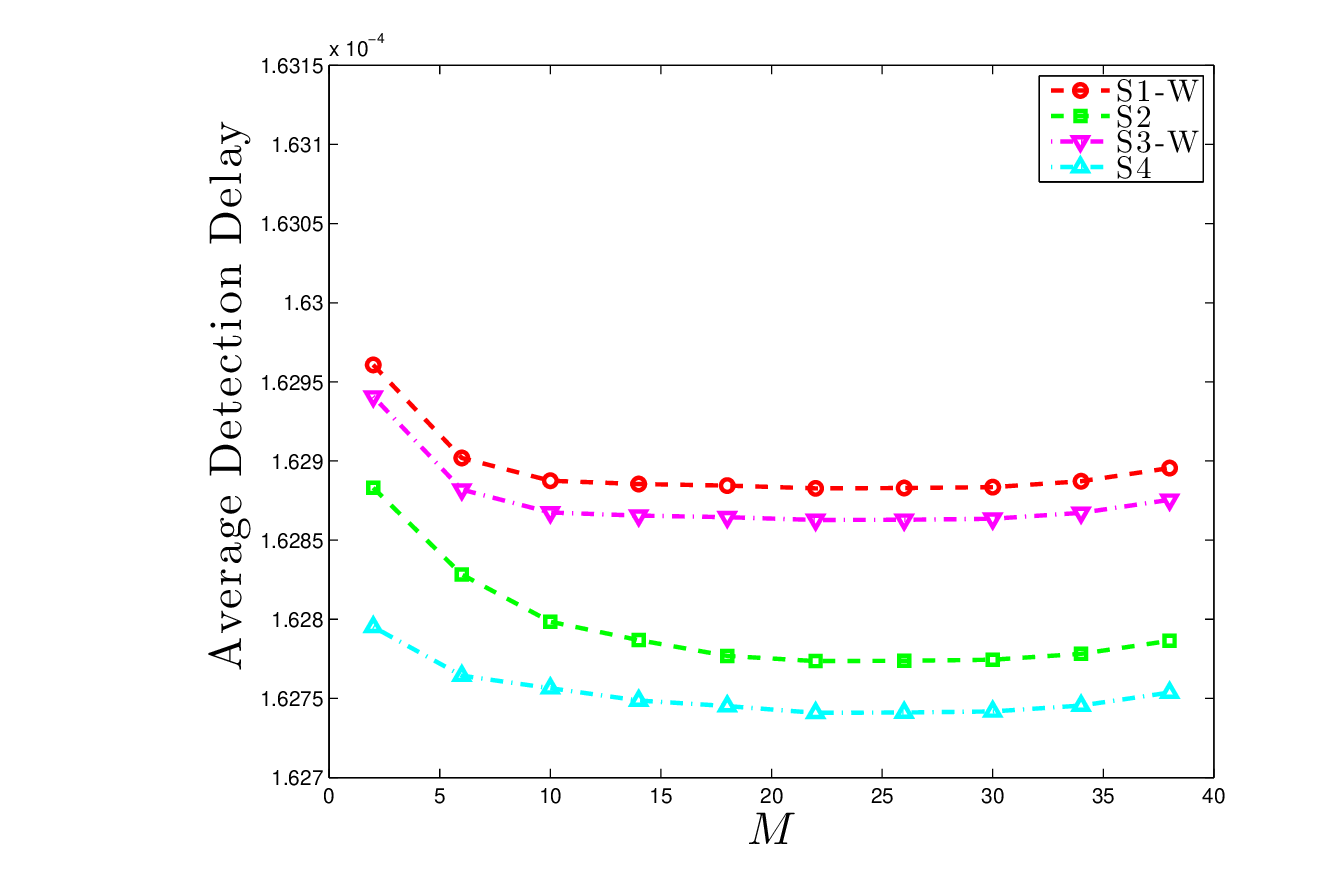}
\caption{Average detection delay vs. the number of transmitters $M$ for the decentralized schemes proposed for the four Swerling target models.}
\label{fig:Sall_M}
\end{figure}

The performances of the decentralized schemes proposed for different target models are compared in Fig. \ref{fig:Sall_M} as $M$ increases. Since we have SNR$=3$dB, the performances of the fast fluctuating (second and fourth) models are superior to those of the slow fluctuating (first and third) models for small $M$ in accordance with Fig. \ref{fig:Sall_snr}. The average LLR increases with increasing $M$ up to some point, similar to the high SNR case, as noted before. Hence, the performance gap between the fast and the slow fluctuating models increases as in Fig. \ref{fig:Sall_snr}. As $M$ further increases the newly added transmitters lie far away from the target, decreasing the average LLR, hence the performances of all schemes degrade. We again observe the advantage of having nonzero-mean channel coefficients as a result of a large scatterer when comparing the first and second models with the third and fourth models, respectively.

\section{Conclusions}

We have considered the sequential distributed detection problem and proposed a novel approach based on level-triggered sampling for energy-constrained wireless sensor networks. Transmitting a single pulse per sample the proposed method enables considerable energy saving. We have analyzed its asymptotic average detection delay performance as its error probabilities tend to zero.
We have then applied the proposed scheme to the decentralized target detection problem by deriving the corresponding detectors under the four Swerling fluctuating target models. Finally, we have provided simulation results that compare the average detection delay performances of the proposed schemes and the conventional decision fusion technique with the majority rule under the four target models in different scenarios. It is seen that the proposed schemes significantly outperform the conventional decision fusion techniques.

\section*{Appendix: Proof of Theorem \ref{thm:finite}}

In the proposed detector, we use the same randomized quantizer as in \cite{Yilmaz12}. The randomization probability $p$, given in \eqref{eq:quant_finite}, as shown in \cite[Lemma 4]{Yilmaz12}, ensures that the likelihood ratio approximations $\exp\left(\hat{L}_{\tilde{\tau}_m}\right)$ and $\exp\left(-\hat{L}_{\tilde{\tau}_m}\right)$ are supermartingales in $m$ with respect to the probability measures $\Pro_0$ and $\Pro_1$, respectively, where the two measures also account for the quantizer randomizations. This is the key property that enables the order-2 asymptotic optimality result in \cite{Yilmaz12}.

The sequential distributed detector proposed in this paper, different from the one in \cite{Yilmaz12}, encodes quantized overshoot information in time by transmitting a pulse after a specifically designed transmission delay, that is smaller than a unit time interval. This difference in the transmission scheme provides significant energy savings compared to the sequential distributed detector in \cite{Yilmaz12}, at the cost of some bounded delay in information arrival times at the FC. Since such delays, which constitute the only difference between two schemes, are within a unit time interval, the stopping time here differs from that in \cite{Yilmaz12} by only a small constant, and the asymptotic analysis performed in \cite{Yilmaz12} holds here. In particular, from \cite[Eq. 42]{Yilmaz12}, we can write
\be
\label{eq:asy_thm}
	0\leq \Exp_i[\hat{\cT}]-\Exp_i[\cT] \leq \frac{R ~\theta}{(\Exp_i[L_1])^2} \frac{|\log \gamma_i|}{\max\{N-1,1\}} [1+o(1)] + \frac{K}{R \tanh(\Delta/2)} + \frac{K\theta}{|\Exp_i[L_1]|} + O(1) + o(1),
\ee
for $i=0,1$, where $R$ is the average message rate [cf. \eqref{eq:delta}], $\theta$ is the overshoot bound [cf. (A2)], $\gamma_1=\alpha$ is the false alarm probability, $\gamma_0=\beta$ is the mis-detection probability, $L_1$ is the global LLR of a single observation at each sensor [cf. \eqref{eq:LLR3}], $N$ is the number of available slots in a unit time interval (i.e., denotes the time resolution), $K$ is the number of sensors, and $\Delta$ is the sampling threshold [cf. \eqref{eq:samp_time}]. In comparison to \cite[Eq. 42]{Yilmaz12}, here we use the term average message rate $R$ instead of the communication period $T$, where $R=\frac{K}{T}$. We also explicitly write the information number $K I_i$ as $|\Exp_i[L_1]|$. The second and third terms inside the summation in \eqref{eq:asy_thm} follows from the terms $\frac{\Delta}{I_i}$ and $\frac{\phi}{I_i}$ in \cite[Eq. 39]{Yilmaz12}, respectively. We also used \eqref{eq:delta} in writing $\frac{K}{R \tanh(\Delta/2)}$ from $\frac{\Delta}{I_i}$. Note that $\tanh(\Delta/2)\in(0,1)$ for $\Delta>0$. Finally, the $O(1)$ term is due the transmission delays in our scheme.

We start with the order-1 asymptotic optimality result, for which we need to show $\frac{\Exp_i[\hat{\cT}]}{\Exp_i[\cT]}=1+\frac{\Exp_i[\hat{\cT}]-\Exp_i[\cT]}{\Exp_i[\cT]}=1+o(1)$, i.e., $\frac{\Exp_i[\hat{\cT}]-\Exp_i[\cT]}{\Exp_i[\cT]}=o(1)$. Using \eqref{eq:asy_thm} and $\Exp_i[\cT] \geq \frac{|\log \gamma_i|}{|\Exp_i[L_1]|}+o(1)$ from \cite[Lemma 2]{Yilmaz12}, we can write
\be
	\frac{\Exp_i[\hat{\cT}]-\Exp_i[\cT]}{\Exp_i[\cT]} \leq \frac{R}{|\Exp_i[L_1]| \max\{N-1,1\}} + \frac{|\Exp_i[L_1]|}{R |\log \gamma_i|} + o(1),
\ee
as $\alpha,\beta\to0$. Then, the first result in Theorem \ref{thm:finite} follows for $|\Exp_i[L_1]|<\infty$.

The order-2 asymptotic optimality result follows directly from \eqref{eq:asy_thm} provided that $|\Exp_i[L_1]|\not=0$ and $R$ is a constant.

\ignore{
We use the following definition of asymptotic optimality.

\begin{dfn}
  A sequential scheme $(\hat{\cT},\hat{\delta}_{\hat{\cT}})$, with stopping time $\hat{\cT}$ and decision function $\hat{\delta}_{\hat{\cT}}$, satisfying the two error probability constraints $\Pro_0(\hat{\delta}_{\hat{\cT}}=1)\leq\alpha$ and $\Pro_1(\hat{\delta}_{\hat{\cT}}=0)\leq\beta$ is said to be order-1 asymptotically optimal if
  \be
  \frac{\Exp_i[\hat{\cT}]} {\Exp_i[\cT]} = 1+o(1)~~\text{as}~\alpha,\beta\to0,
  \ee
  where $\cT$ is the stopping time of the optimal SPRT that satisfies the error probability constraints with equality and $o(1)$ denotes a vanishing term as $\alpha,\beta\to0$. It is order-2 asymptotically optimal if
  \be
  \Exp_i[\hat{\cT}] - \Exp_i[\cT] = O(1),
  \ee
  where $O(1)$ denotes a constant term, and order-3 asymptotically optimal if
  \be
  \Exp_i[\hat{\cT}] - \Exp_i[\cT] = o(1).
  \ee
\end{dfn}

Note that a higher order implies a stronger type of asymptotic optimality, i.e., order-3 $\Rightarrow$ order-2 $\Rightarrow$ order-1, where $\Rightarrow$ means ``implies". Order-1 asymptotic optimality is the most commonly used type in the literature, but also the weakest type. The average detection delay of an order-1 scheme can diverge from that of the optimal scheme. On the other hand, under order-2 and order-3 optimality the average detection delay remains parallel and converges to that of the optimal scheme, respectively. Order-3 asymptotic optimality is extremely rare in the sequential analysis literature, and corresponds to schemes that are considered as optimum \emph{per se} for all practical purposes.

Next we show that the proposed scheme with infinite time resolution is order-2 asymptotically optimal.
It was shown in \cite[Theorem 2]{Yilmaz13} that the asymptotic expressions for average detection delays $\Exp_i[\hat{\cT}], i=0,1$ of a decentralized detection scheme based on the level-triggered sampling, e.g., the one described in Section \ref{sec:time_enc}, are in general given by
  \begin{align}
  \label{eq:asy_delay_gen}
        \Exp_1[\hat{\cT}]=&\frac{|\log \alpha|}{\hat{I}_1(1)}+O(1) ~~
        \text{and}~~\Exp_0[\hat{\cT}]=\frac{|\log \beta|}{\hat{I}_0(1)}+O(1),
  \end{align}
  as $\alpha,\beta\to0$, where $\hat{I}_i(1) \triangleq \sum_{k=1}^K \frac{\hat{I}_i^k(t_1^k)}{I_i^k(t_1^k)}I_i^k(1), i=0,1$ are hypothetical Kullback-Leibler (KL) information numbers defined for analytical purposes; $I_i^k(1)\triangleq|\Exp_i[l_1^k]|$, $I_i^k(t_1^k)\triangleq|\Exp_i[\lambda_1^k]|$, $\hat{I}_i^k(t_1^k)\triangleq|\Exp_i[\hat{\lambda}_1^k]|, i=0,1$ are the KL information numbers of the LLR sequences $\{l_t^k\}_t$, $\{\lambda_n^k\}_n$, $\{\hat{\lambda}_n^k\}_n$, respectively. Here, at sensor $k$ and under $\Hyp_i$, $\Exp_i[l_1^k]$ is the average LLR of a single observation [cf. \eqref{eq:LLR3}]; $\Exp_i[\lambda_1^k]$ is the average LLR accumulated in a local SPRT (i.e., during a sampling interval); and $\Exp_i[\hat{\lambda}_1^k]$ is the average LLR transmitted in a message. Hence, the hypothetical KL information number $\hat{I}_i(1)$ is, in fact, the projection of $I_i(1)=\sum_{k=1}^K I_i^k(1)$, which is the KL information number of the global LLR sequence $\{\sum_{k=1}^K l_t^k\}$, onto the filtration generated by the transmitted bit sequence. Since sensors do not transmit the LLR of each single observation (they transmit LLR messages of several observations), $\hat{I}_i(1)$ is not a real KL information number, but it is a projected or hypothetical KL information number, which plays a key role in the asymptotic analysis on the average detection delay \cite{Yilmaz13}. If the random overshoot $q_n^k$ is not fully encoded in each transmitted message, then this projected KL information number $\hat{I}_i(1)$, as in \cite{Fellouris11,Yilmaz12,Yilmaz13}, will not be the same as the original KL information number $I_i(1)$ since the average LLR in a transmitted message will be different from the average LLR accumulated in a local SPRT, i.e., $\hat{I}_i^k(t_1^k)\not=I_i^k(t_1^k)$. However, for our proposed scheme we have $\hat{I}_i^k(t_1^k)=I_i^k(t_1^k)$ since we encode the overshoot $q_n^k$ into the transmission delay $\xi_n^k$, as explained in Section \ref{sec:time_enc}. Therefore, from \eqref{eq:asy_delay_gen} we write the asymptotic average detection delays of the proposed scheme as
  \begin{align}
  \label{eq:asy_delay}
        \Exp_1[\hat{\cT}]=&\frac{|\log \alpha|}{I_1(1)}+O(1) ~~
        \text{and}~~\Exp_0[\hat{\cT}]=\frac{|\log \beta|}{I_0(1)}+O(1),
  \end{align}
  as $\alpha,\beta\to0$.

  For the optimal centralized scheme we have, from \cite[Lemma 2]{Yilmaz12}, $\Exp_1[\cT]\geq\frac{|\log \alpha|}{I_1(1)}+o(1)$ and $\Exp_0[\cT]\geq\frac{|\log \beta|}{I_0(1)}+o(1)$, which is, together with \eqref{eq:asy_delay}, sufficient to prove order-2 asymptotic optimality, i.e, \eqref{eq:thm}. In order to show that we cannot achieve order-3 asymptotic optimality, following \cite[Theorem 2]{Yilmaz13} we can write $\Exp_1[\cT]=\frac{|\log \alpha|}{I_1(1)}+O(1)$ and $\Exp_0[\cT]=\frac{|\log \beta|}{I_0(1)}+O(1)$, where the $O(1)$ term includes the average excess LLR in $L_{\cT}$ over $A$ or below $-B$. Note that the excess LLR here is produced by the global LLR $\sum_{k=1}^K l_{\cT}^k$ of the single observations at the stopping time $\cT$. On the other hand, the average excess LLR included in the $O(1)$ term in \eqref{eq:asy_delay} is a consequence of the final LLR message(s) received at the stopping time $\hat{\cT}$ from any sensor(s). Obviously, the average excess LLR produced by message(s), including the LLR of several observations, is larger than that produced by a single observation. Hence, the difference $\Exp_i[\hat{\cT}]-\Exp_i[\cT]$ remains as a bounded constant.
}


\end{document}